\documentclass[11pt]{article}
\usepackage{amsmath}
 \numberwithin{equation}{section}
 \allowdisplaybreaks
 \newcommand{\be}{\begin{equation}}
 \newcommand{\ee}{\end{equation}}
 \newcommand{\bea}{\begin{eqnarray}}
 \newcommand{\eea}{\end{eqnarray}}
 \newcommand{\ds}{\displaystyle}
 \newcommand{\nn}{\nonumber}
 \newcommand{\td}{\tilde}
 \newcommand{\wtd}{\widetilde}
 \newcommand{\pd}{\partial}
 \newcommand{\one}{{\bf 1}}

 \newcommand{\bk}{{\bf k}}

 \newcommand{\bw}{{\bf w}}
 
 \newcommand{\bJ}{{\bf J}}
 \newcommand{\bL}{{\bf L}}
 \newcommand{\bM}{{\bf M}}
 \newcommand{\bQ}{{\bf Q}}
 \newcommand{\cA}{{\cal A}}

 \newcommand{\cH}{{\cal H}}
 
 \newcommand{\cL}{{\cal L}}
 \newcommand{\cM}{{\cal M}}
 
 \newcommand{\cO}{{\cal O}}

 \newcommand{\dtp}{\dot{p}}
 \newcommand{\dtq}{\dot{q}}
\newcommand{\tamphys}{\it George P. \& Cynthia Woods Mitchell  Institute
for Fundamental Physics and Astronomy,\\
Texas A\&M University, College Station, TX 77843, USA}

\newcommand{\auth}{Jianwei Mei}
\begin{document}

\begin{flushright}
\hfill{
MIFP-10-04}\\
%\hfill{
%\bf hep-th/yymmnnn}
\end{flushright}

\begin{center}

{\large\bf The Entropy for General Extremal Black Holes}

\vspace{25pt}

\auth

\vspace{10pt}{\tamphys}

\vspace{25pt}

\underline{ABSTRACT}

\end{center}

We use the Kerr/CFT correspondence to calculate the entropy for
all known extremal stationary and axisymmetric black holes. This
is done with the help of two ansatzs that are general enough to
cover all such known solutions. Considering only the contribution
from the Einstein-Hilbert action to the central charge(s), we find
that the entropy obtained by using Cardy's formula exactly matches
with the Bekenstein-Hawking entropy.

\vspace{15pt}

\thispagestyle{empty}

\pagebreak
\setcounter{page}{1}

\tableofcontents

%\newpage
\section{Introduction}

To successfully calculate the entropy for black holes is a
challenge for all candidates of the quantum gravity theory. In
reverse, helpful insight to quantum gravity may be obtained if one
can find a general way to calculate the black hole entropy.

The Kerr/CFT correspondence \cite{ghss08,hmns08} has been quite
successful with calculating the entropy for extremal black holes.
The basic idea is to discuss dynamics on the near-horizon metric
of the black holes. With appropriate boundary conditions, the
corresponding phase space can be identified with that of a two
dimensional conformal field theory. The entropy of the black hole
can then be calculated from the corresponding central charge(s) by
using Cardy's formula. After it was first proposed in
\cite{ghss08}, the method has been found to work for all the cases
that have been checked (for refs, see \cite{hss09}). It was
suggested in \cite{hmns08} that the extremal condition may be at
the heart of the correspondence. So the Kerr/CFT correspondence is
also called the Extremal Black Hole/CFT correspondence.

In hindsight, several important points have also been raised in
\cite{hmns08}. The first is related to the matter field
contribution to the central charges of the dual CFTs. It was found
in \cite{hmns08} that the gauge field does not contribute to the
central charge for solutions in the Einstein-Maxwell system in
four dimensions. This result was echoed in
\cite{ghezelbash09,cmn09}, where it was shown by using examples in
four and five dimensions that non-gravitational fields such as the
scalar field, the Abelian gauge field and the antisymmetric tensor
field do not contribute to the central charge(s). The second point
is that the success of the Kerr/CFT calculation may partially due
to the possibility that all near-horizon metrics share a
particular common structure. The near-horizon metrics for some
extremal black holes have been studied in
\cite{kunduri.lucietti.reall07,figueras.kunduri.lucietti.rangamani08}
in a different context. In four dimensions, the near-horizon
metrics are found to be of the form
%%%
\be ds_4^2=f(\theta)\Big[-r^2dt^2+\frac{dr^2}{r^2} +\alpha
(\theta)d\theta^2\Big]+\gamma(\theta)(d\phi+k r dt)^2\,,\ee
%%%
while in higher dimensions they are found to be of the form
%%%
\be ds_d^2=f(\theta^i)\Big[-r^2dt^2+\frac{dr^2}{r^2}\Big]
+\alpha_i(\theta^j) d\theta^{i2}+\gamma_{ab}(\theta^i)(d\phi^a
+k_a r dt)(d\phi^b+k_b r dt) \label{metric.near.horizon}\ee
%%%
for a certain class of solutions, where $k$ and $k_a$ are
constants while all the functions depend on $\theta^i$'s only. It
was then shown in \cite{chow.cvetic.lu.pope08} that
(\ref{metric.near.horizon}) indeed plays a significant role when
the Kerr/CFT correspondence is applied to various solutions in
(gauged) supergravity theories. Further examples were also
presented in \cite{lu.mei.pope.vazquez-poritz09}. Lastly, it was
speculated in \cite{hmns08} that the Frolov-Thorne temperature may
be of the general form $T_L=\frac1{2\pi k}$ in four dimensions.
This was then generalized to higher dimensions in
\cite{chow.cvetic.lu.pope08},
%%%
\be T^a_L=\frac1{2\pi k_a}\,,\label{def.FT.temp}\ee
%%%
based on all the examples that have been studied. This result also
plays a crucial role in applying the Kerr/CFT correspondence to
various black hole solutions
\cite{chow.cvetic.lu.pope08,lu.mei.pope.vazquez-poritz09}.

In this paper, we present two ansatzs that are general enough to
cover all known stationary and axisymmetric black holes. Extra
constraints can be obtained by noticing that black hole horizons
are intrinsically regular. We then show that
(\ref{metric.near.horizon}) can be derived as soon as the
near-horizon limit is taken for extremal black holes. As a result,
(\ref{metric.near.horizon}) is valid for all known extremal
stationary and axisymmetric black holes. The Frolov-Thorne
temperature of the form (\ref{def.FT.temp}) is also derived in a
straight forward manor. Then we explicitly calculate the central
charge(s) related to (\ref{metric.near.horizon}). When the
microscopic entropy is calculated by using Cardy's formula, we
find that the result exactly matches with the Bekenstein-Hawking
entropy. In this way, we demonstrate in a general fashion that the
Kerr/CFT correspondence is applicable to all known extremal
stationary and axisymmetric black holes. What's more, empirical
results such as (\ref{def.FT.temp}) can also be derived without
making extra assumptions.

Note earlier works have largely demonstrated the general
applicability of the Kerr/CFT correspondence (see, e.g.
\cite{lu.mei.pope08c,chow.cvetic.lu.pope08,lu.mei.pope.vazquez-poritz09}).
So it is not our intention here to show this again. Rather, we are
most interested to see to what extent can the calculation be
carried out in a general fashion.

For practical reasons, we have only considered the contribution
from the Einstein-Hilbert action to the central charge(s). The
fact that the resulted microscopic entropy matches with the
Bekenstein-Hawking entropy implies that the non-gravitational
contributions to the central charge(s) are zero, which is
consistent with the results found in
\cite{hmns08,ghezelbash09,cmn09}. One can certainly try to repeat
the same process for more complicated theories. For example, it
has been shown in \cite{acott09} (See also
\cite{krishnan.kuperstein09} for an earlier work) that in a theory
with higher-derivative corrections in the gravitational sector,
the higher-derivative terms also contribute to the central
charge(s) and the correct entropy is the one constructed by Iyer
and Wald \cite{wald93,iyer.wald94}. However, it is obvious that a
similar calculation will be extremely difficult.

The plan of the paper is as following. In section
\ref{sec.ansatz}, we will present the two ansatzs for all known
stationary and axisymmetric black holes. The near-horizon metric
for extremal black holes will then be derived in section
\ref{sec.NHEM}. The central charges will be calculated in section
\ref{sec.central}, but most of the extra detail will be contained
in Appendix \ref{app.central}. The microscopic entropy from the
CFT side is then calculated in section \ref{sec.entropy}. A
summary will be given in section \ref{sec.summary}.

To make the whole calculation more accessible to most readers, we
have included an introduction to the treatment of asymptotic
symmetries by using the covariance phase space method in Appendix
\ref{app.symmetry}. We will also revisit most of the examples
studied in
\cite{lu.mei.pope08c,chow.cvetic.lu.pope08,lu.mei.pope.vazquez-poritz09}
in Appendix \ref{app.examples}, by using the new perspective that
we gain from the present work.

\section{Two General Ansatzs for Stationary and Axisymmetric Black Holes}
\label{sec.ansatz}

The basics of the Kerr/CFT correspondence has been explained in
\cite{ghss08} in much detail. Here we will go directly to the
general case we want to study.

We will start with presenting two general ansatzs that cover all
known stationary and axisymmetric black hole solutions. The
construction will be partially based on our experience with all
the solutions that are known.

Stationary and axisymmetric black hole solutions share some common
features:
%%%
\begin{itemize}
\item By using the term ``stationary and axisymmetric", one
assumes that (i) a coordinate system exists where some of the
coordinates can be identified with the asymptotic time direction
$\hat t$ and the azimuthal directions $\hat\phi^a$, and (ii) the
metric does not depend on $\hat t$ nor $\hat\phi^a$.
\item Among the rest of the coordinates, one coordinate can be
singled out as describing the radial direction $\hat r$. For all
known solutions, the position of the black hole horizon ($\hat
r=r_H$) is determined by a single function of $\hat r$ :
$\Delta(r_H)=0$.
\item All other coordinates are then related to the latitudinal
angles $\theta^i$. For a black hole in $d$-dimensional spacetime,
there can be $[\frac{d-1}{2}]$ independent rotations. So
$a=1,\cdots,[\frac{d-1}{2}]$ and $i=1,\cdots,[\frac{d}{2}]-1$.
\item For all known solutions, one can always chose the coordinate
systems so that the metrics do not have any cross terms involving
$d\hat r$ or $d\theta^i$.
\item Near the black hole horizon, it can either be a term like
$d\hat t +f_a(\hat r,\theta^i) d\hat\phi^a$ or a term like
$f_a(\hat r,\theta^i) d\hat\phi^a$ playing the role of time.
\end{itemize}
%%%
Metrics reflecting such features can always be written as
%%%
\be ds_d^2=-\frac{\Delta}{f_t}\Big[d\hat t +f_a d\hat\phi^a\Big]^2
+\frac{f_r}{\Delta}d\hat r^2+g_{ij}d\theta^i d\theta^j
+d\bar{s}_\phi^2\,,\label{metric.general.a1}\ee
%%%
or
%%%
\be ds_d^2=-\frac{\Delta}{f_t}\Big[f_a d\hat\phi^a\Big]^2
+\frac{f_r}{\Delta}d\hat r^2+g_{ij}d\theta^i d\theta^j
+d\bar{s}_\phi^2\,,\label{metric.general.a2}\ee
%%%
with
%%%
\be d\bar{s}_\phi^2=g_{ab}(d \hat\phi^a-\chi_ad\hat t)
(d\hat\phi^b -\chi_b d\hat t)+f_{tt} d\hat t^2\,.
\label{metric.general.b}\ee
%%%
Note all the functions depend on $\hat r$ and $\theta^i$'s only,
while $\Delta$ will be the function determining the location of
the horizon and so it depends on $\hat r$ only. We have allowed
$d\theta^i$'s to mix among themselves in (\ref{metric.general.a1})
and (\ref{metric.general.a2}), so both ansatzs can describe
possibly slightly more general cases than listed above. We have
also included the $f_{tt}d\hat t^2$ term in
(\ref{metric.general.b}) to make (\ref{metric.general.a1}) and
(\ref{metric.general.a2}) as general as possible. The assumption
on $f_{tt}$ is that it should not play any significant role near
the horizon. As we will see below, this means $f_{tt}\sim
\Delta^2$ as $\hat r\rightarrow r_H$. As far as we can tell, all
known stationary and axisymmetric black holes can either be
written in the form of (\ref{metric.general.a1}) or in the form of
(\ref{metric.general.a2}). We also notice that the two ansatz are
actually general enough to go beyond black holes and cover objects
such as the black ring \cite{emparan.reall01}.

Some extra constraints can be obtained for the functions in
(\ref{metric.general.a1}), (\ref{metric.general.a2}) and
(\ref{metric.general.b}) by noticing that black hole horizons are
intrinsically regular. A regular horizon means that the metric
(and the matter fields) should be manifestly regular on the
horizon if the coordinate system is chosen appropriately.

To see how this can help us, note that the first two terms in
(\ref{metric.general.a1}) can be written as
%%%
\be\frac{\Delta}{f_t}\left(-\Big[d\hat t +f_a\,d\hat\phi^a\Big]^2
+\frac{f_tf_r}{\Delta^2}d\hat r^2\right)=-\frac{\Delta}{f_t}\cA^2
+2\sqrt{f_r/f_t}\;d\hat r\cA\,,\label{metric.feature1}\ee
%%%
where
%%%
\be\cA=d\hat t +f_a\,d\hat\phi^a+\frac{\sqrt{f_tf_r}}{\Delta}d\hat
r\,. \label{metric.feature.def.A}\ee
%%%
The superficial singularity near the horizon comes solely from
$\Delta(r_H)=0$. To make the metric regular on the horizon, one
can try to make $\cA$ regular first. This can be achieved if there
exist functions $h_v=h_v(\hat r)$, $h_a=h_a(\hat r)$ and $h_\cA
=h_\cA(\hat r, \theta^i)$ being regular on the horizon and
satisfying
%%%
\be\sqrt{f_tf_r}=h_v +f_a h_a +h_\cA\Delta+\cO(\Delta^2)\,.
\label{metric.feature3a}\ee
%%%
In this case one can write $\cA=dv +f_a\,d\psi^a +h_\cA d\hat r
+\cO(\Delta)$ by using the coordinate transformation
%%%
\be dv=d\hat t +\frac{h_v(\hat r)}{\Delta(\hat r)}d\hat r\,,\quad
d\psi^a=d\hat\phi^a +\frac{h_a(\hat r)}{\Delta(\hat r)}d\hat r\,.
\label{metric.feature2}\ee
%%%
We find that this process is possible for all know examples. For
(\ref{metric.general.b}),
%%%
\bea d\bar{s}_\phi^2&=&g_{ab}\Big(d\psi^a-\chi_adv -\frac{h_a
-\chi_a h_v}{\Delta}d\hat r\Big)\Big(d\psi^b-\chi_bdv -\frac{h_b
-\chi_b h_v}{\Delta}d\hat r\Big)\nn\\
&&+f_{tt}\Big(dv-\frac{h_v}{\Delta}d\hat r\Big)^2\,.\eea
%%%
To make $d\bar{s}_\phi^2$ regular on the horizon, one must have
%%%
\be\chi_a=\frac{h_a+h_\chi^a\Delta}{h_v} +\cO(\Delta^2) \,,\quad
f_{tt}=h_{tt}\Delta^2 +\cO(\Delta^3)\,.\ee
%%%
Again $h_\chi^a=h_\chi^a(\hat r,\theta^i)$ and $h_{tt}=h_{tt}
(\hat r,\theta^i)$ must be regular on the horizon. Using these
results and keeping only leading order corrections, one has for
(\ref{metric.general.a1}) at $\hat r\rightarrow r_H$,
%%%
\bea ds_d^2&\approx&f_r\Big\{-\Delta\frac{(d\hat t +f_a
d\hat\phi^a)^2}{ (h_v +f_a h_a +h_\cA\Delta)^2} +\frac{d\hat
r^2}{\Delta}\Big\} +g_{ij}d\theta^i d\theta^j +h_{tt}
\Delta^2d\hat t^2\nn\\
&&\quad+g_{ab}\Big(d \hat\phi^a-\frac{h_a+h_\chi^a \Delta}{h_v}
d\hat t\Big)\Big(d\hat\phi^b-\frac{h_b+h_\chi^b\Delta}{h_v}d\hat
t\Big) \,.\label{metric.general.c}\eea
%%%
If the same process is repeated for (\ref{metric.general.a2}), one
can find that when $\hat r\rightarrow r_H$,
%%%
\bea ds_d^2&\approx&f_r\Big\{-\Delta\frac{(f_a
d\hat\phi^a)^2}{(f_a h_a +h_\cA\Delta)^2}+\frac{d\hat
r^2}{\Delta}\Big\}+g_{ij}d\theta^i
d\theta^j+h_{tt}\Delta^2d\hat t^2\nn\\
&&\quad+g_{ab}\Big(d
\hat\phi^a-\frac{h_a+h_\chi^a\Delta}{h_v}d\hat t\Big)
\Big(d\hat\phi^b -\frac{h_b+h_\chi^b\Delta}{h_v}d\hat t\Big)
\,.\label{metric.general.d}\eea
%%%
As we will show in Appendix \ref{app.examples},
(\ref{metric.general.c}) with $h_\cA=h_\chi^a=h_{tt}=0$ is in fact
exact (i.e., not an approximation) for a surprisingly large number
of solutions.

Strictly speaking, our derivation of (\ref{metric.general.c}) and
(\ref{metric.general.d}) is by no means the most general one. The
whole process rests upon using the coordinate transformation
(\ref{metric.feature2}) to render both $\cA$ and $d\bar{s}_\phi^2$
finite on the horizon {\it separately}. One may as well try to
think of other ways to make the whole metric
(\ref{metric.general.a1}) finite on the horizon all together.
Since we have made no effort trying in such a direction, we will
have nothing to say about this point. For the purpose of the
paper, it is important to notice that (\ref{metric.general.c}) and
(\ref{metric.general.d}) already appear to be general enough to
cover all known stationary and axisymmetric black hole solutions.

For later convenience, lets calculate the black hole temperature
for (\ref{metric.general.c}) and (\ref{metric.general.d}). For
that purpose, we choose a static coordinate system with both $\hat
t$ and $\hat\phi^a$ canonically normalized. The surface gravity is
calculated with the particular Killing vector,
%%%
\be\xi=\pd_{\hat t}+\Omega_a\pd_{\hat\phi^a}\,.
\label{killing.major}\ee
%%%
Here the constants $\Omega_a$'s are chosen to make $\xi$ null on
the (outer) horizon. They are interpreted as the angular
velocities corresponding to the azimuthal angles $\hat\phi^a$. To
see how $\Omega_a$'s can be calculated, note that for
(\ref{metric.general.c}),
%%%
\be\xi^2=\frac{-f_r\Delta\cdot(1 +f_a \Omega_a)^2}{(h_v +f_a h_a
+h_\cA\Delta)^2} +g_{ab}\Big(\Omega_a-\frac{h_a +h_\chi^a \Delta}{
h_v} \Big)\Big(\Omega_b -\frac{h_b+h_\chi^b \Delta}{h_v}\Big)
+h_{tt}\Delta^2\,,\label{killing.xi2}\ee
%%%
and for (\ref{metric.general.d}),
%%%
\be\xi^2=\frac{-f_r\Delta\cdot(f_a \Omega_a)^2}{(f_a h_a +h_\cA
\Delta)^2} +g_{ab}\Big(\Omega_a-\frac{h_a +h_\chi^a \Delta}{
h_v}\Big)\Big(\Omega_b -\frac{h_b+h_\chi^b \Delta}{h_v}\Big)
+h_{tt}\Delta^2\,.\ee
%%%
For both cases, to make $\xi$ vanish on the horizon one must have
%%%
\be\Omega_a=\frac{h_a^0}{h_v^0}\,,\quad h_a^0=h_a(r_H)\,,\quad
h_v^0=h_v(r_H)\,.\label{def.omega.a}\ee
%%%
Including corrections to the leading order, one has
%%%
\be \frac{h_a}{h_v}=\Omega_a+\Omega_a'\cdot(\hat r-r_H) +\cO(\hat
r-r_H)^2 \,,\quad \Omega_a'\equiv\Big(\frac{h_a}{h_v}
\Big)'\Big|_{\hat r=r_H} \,.\label{def.omega.a.correction}\ee
%%%
The surface gravity on the horizon can be calculated by using
%%%
\be\kappa^2=\frac{(\pd\lambda)^2}{4\lambda}\Big|_{\hat r=r_H}
\,,\quad\lambda=-\xi^2\,. \label{def.kappa2}\ee
%%%
For non-extremal solutions, $\Delta(\hat r)=\Delta_0'\cdot(\hat
r-r_H) +\cO(\hat r-r_H)^2$ with $\Delta_0'=\Delta'(r_H)$. So to
leading order,
%%%
\be\lambda=\frac{f_r^0}{h_v^{02}}\Delta_0'\cdot(\hat r-r_H)
+\cO(\hat r-r_H)^2\,,\ee
%%%
where $f_r^0=f_r(r_H,\theta^i)$. The surface gravity
(\ref{def.kappa2}) is then given by
%%%
\be\kappa^2=\frac{g^{rr}\pd_{\hat r}\lambda\pd_{\hat r} \lambda}{4
\lambda} \Big|_H =\frac{\Delta_0'^2}{4h_v^{02}}\,.\ee
%%%
So the temperature of the black hole is given by
%%%
\be T_H=\frac{\kappa}{2\pi} =\frac{\Delta_0'}{4\pi h_v^0}
\,.\label{def.T.nonextremal}\ee
%%%
For an extremal solution, $\Delta=\frac12\Delta_0''\cdot(\hat
r-r_H)^2 +\cO(\hat r-r_H)^3$ with $\Delta_0''=\Delta''(r_H)$. One
can find that $T_H=0$. An easy way to see this is to start from
(\ref{def.T.nonextremal}) and then take the extremal limit
%%%
\be\Delta_0'\rightarrow0\quad\Longrightarrow\quad
T_H\rightarrow0\,.\ee
%%%
Note all the results starting from (\ref{def.omega.a}) are valid
for both (\ref{metric.general.c}) and (\ref{metric.general.d}).

\section{The Near-Horizon Metric for Extremal Black Holes}
\label{sec.NHEM}

To get the near-horizon metric for an extremal black hole, one
follows \cite{bardeen.horowitz99,ghss08,chow.cvetic.lu.pope08} and
let
%%%
\be \hat r=r_H+y\lambda\,r_H\,,\quad \hat t=\frac{2 h_v^0}{
\lambda\, r_H\Delta_0''}\td{t}\,,\quad \hat\phi^a=\phi^a
+\Omega_a\hat t\,. \label{etremal.limit}\ee
%%%
Using $\Delta=\frac12\Delta_0''\cdot(\hat r-r_H)^2 +\cO(\hat
r-r_H)^3$ and after sending $\lambda\rightarrow0$, one has for
both (\ref{metric.general.c}) and (\ref{metric.general.d}),
%%%
\bea ds^2&=&\frac{2f_r^0}{\Delta_0''}\Big(-y^2d\td{t}^2
+\frac{dy^2}{y^2}\Big)+g_{ij}^0d\theta^i d\theta^j\nn\\
&&+g_{ab}^0(d\phi^a+k^ayd\td{t})(d\phi^b+k^byd\td{t})\,,
\label{metric.general.nhem1}\eea
%%%
where $g_{ij}^0=g_{ij}(r_H,\theta^i)$, and we have used
(\ref{def.omega.a.correction}) and have defined
%%%
\be k^a=-\frac{2h_v^0\Omega_a'}{\Delta_0''}\,.
\label{cft.def.ka}\ee
%%%
One can see that (\ref{metric.general.nhem1}) is exactly of the
form (\ref{metric.near.horizon}). Based on the argument made in
the previous section, (\ref{metric.general.nhem1}) is valid for
all extremal stationary and axisymmetric black holes.

To get to the global coordinates, let
%%%
\be y=r+\sqrt{1+r^2}\cos t\,,\quad \td{t}=\frac{\sqrt{1+r^2} \sin
t}{y}\,.\ee
%%%
Then
%%%
\bea -y^2d\td{t}^2+\frac{dy^2}{y^2}=-(1+r^2)dt^2+\frac{
dr^2}{1+r^2}\,,\nn\\
yd\td{t}=rdt+d\ln\Big(\frac{1+\sqrt{1+r^2}\sin t}{ \cos t+r\sin
t}\Big)\,.\eea
%%%
So by letting
%%%
\be\phi^a\rightarrow\phi^a-k^a\ln\Big(\frac{1+\sqrt{1+r^2}\sin t}{
\cos t+r\sin t}\Big)\,,\ee
%%%
one can rewrite the near-horizon metric
(\ref{metric.general.nhem1}) as
%%%
\bea ds^2&=&\frac{2f_r^0}{\Delta_0''}\Big[-(1+r^2)dt^2
+\frac{dr^2}{1+r^2}\Big]+g_{ij}^0d\theta^i d\theta^j\nn\\
&&+g_{ab}^0(d\phi^a+k^ardt)(d\phi^b+k^brdt)\,.
\label{metric.general.nhem2}\eea
%%%
The significance of this form of the near-horizon metric in the
context of the Kerr/CFT correspondence was first noticed in
\cite{hmns08}, then the importance was stressed upon again in
\cite{chow.cvetic.lu.pope08} for black hole solutions in higher
dimensions. More examples were then provided in
\cite{lu.mei.pope.vazquez-poritz09}.

\section{The Central Charge(s) of the Dual CFT(s)}
\label{sec.central}

Following \cite{ghss08} one can try to calculate the black hole
entropy by studying dynamics on the near-horizon metric
(\ref{metric.general.nhem2}), with the help of appropriate
boundary conditions. The symmetries of the corresponding phase
space are generated by $[\frac{d-1}{2}]$ commuting generators
\cite{chow.cvetic.lu.pope08}, namely
%%%
\be \xi^a_m=-e^{-im\phi^a}\,\pd_{\phi^a}-imre^{-im\phi^a}
\pd_r\,,\quad a=1\,,\cdots\,,[\frac{d-1}{2}]\,.
\label{def.classical.generators}\ee
%%%
It is easy to check that
%%%
\be i[\xi_m^a\,,\,\xi_n^a]=(m-n)\xi_{m+n}^a\,.\ee
%%%
These transformations generate $[\frac{d-1}{2}]$ commuting
Virasoro algebras. For each Virasoro algebra, the phase space can
be identified with that of a two-dimensional conformal field
theory. The classical version of the charge $Q_{\xi_m^a}$ is
defined in (\ref{cov.def.Qxi}). To get the quantum version of the
charge, we write
%%%
\be Q_{\xi_m^a}=L_m^a-\alpha\delta_m\,,
\label{cft.def.quantum.operator}\ee
%%%
with $\alpha$ being some constant. From (\ref{cov.def.Qxi}) and
(\ref{cov.def.kxi.GR}), it is easy to see that if $\xi_m^a$ is
scaled by a factor, the right hand side of
(\ref{cft.def.quantum.operator}) also needs to be scaled by the
same factor. Especially, one has
%%%
\be Q_{[\xi_m^a,\xi_n^a]}=Q_{-i(m-n)\xi_{m+n}^a} =-i(m-n)
\Big(L_{m+n}^a-\alpha\delta_{m+n}\Big)\,.\ee
%%%
So from (\ref{cov.def.PB.general}),
%%%
\bea[L_m^a\,,\,L_n^a]&=&i\Big\{Q_{\xi_m^a}\,,\,Q_{\xi_n^a}\Big\}_{P.B.}
=i\Big(Q_{[\xi_m^a,\xi_n^a]}+K[\xi_m^a,\xi_n^a]\Big)\nn\\
&=&(m-n)L_{m+n}-2m\alpha\delta_{m+n}+i K[\xi_m^a,\xi_n^a]\,.\eea
%%%
Comparing this with the usual relation,
%%%
\be[L_m^a\,,\,L_n^a]=(m-n)L_{m+n}^a +\frac{c^a}{12}m(m^2-1)
\delta_{m+n}\,,\ee
%%%
one gets
%%%
\be K[\xi_m^a,\xi_n^a]=-i\frac{c^a}{12}m\Big(m^2-1 +\frac{24
\alpha}{c^a}\Big)\delta_{m+n}\,.\label{def.central.ca}\ee
%%%
So the central charge $c^a$ is determined by the coefficient of
the $m^3$ term in $K[\xi_m^a,\xi_n^a]$. The term linear in $m$ is
not so important because $\alpha$ is a free parameter.

The central term $K[\xi_m^a,\xi_n^a]$ corresponding to the
near-horizon metric (\ref{metric.general.nhem2}) is calculated in
(\ref{def.value.Kmn}),
%%%
\be K[\xi_m^a,\xi_n^a]=-\frac{i(m-n)n^2 k^a}{16\pi}
\delta_{m+n}\cA_{rea}\,,\ee
%%%
with $\cA_{rea}$ being the horizon area for either
(\ref{metric.general.c}) or (\ref{metric.general.d}). Comparing
this result with (\ref{def.central.ca}), one has
%%%
\be c^a=\frac{3k^a}{2\pi}\cA_{rea}\,.\label{def.ca.general}\ee
%%%
Note this result only contains the contribution from the
Einstein-Hilbert action.

\section{The Entropy}\label{sec.entropy}

In the following, we shall try to relate the central charge to the
entropy by using Cardy's formula. Again following \cite{ghss08},
one can adopt the Frolov-Thorne vacuum \cite{frolov.thorne89} to
provide a definition of the vacuum state for the extremal metric.
One important task here is to derive the left-moving and
right-moving temperatures. We will do it by starting with
non-extremal metrics and then take the extremal limit.

Quantum fields for the general (non-extremal) metrics
(\ref{metric.general.a1}) and (\ref{metric.general.a2}) can be
expanded in eigenstates with asymptotic energy $\omega$ and
angular momentum $m_a$, with $\hat{t}$ and $\hat\phi^a$ dependence
$e^{-i \omega\hat{t}+ i m_a \hat\phi^a}$. In terms of the
redefined $\td{t}$ and $\phi^a$ coordinates of the extremal
near-horizon limit, given by (\ref{etremal.limit}), we have
%%%
\be e^{-i \omega\hat{t}+ i m_a\hat\phi^a} = e^{-i n_R\td{t} +
in_L^a \phi^a}\,,\ee
%%%
with\footnote{From now on until (\ref{cft.def.TL}), any quantity
from the extremal solution will be distinguished with a tilde. For
example, $\td\Omega_a$ is an angular velocity for the extremal
solution, while $\Omega_a$ is its counterpart for the non-extremal
solution.}
%%%
\be n_L^a=m_a\,,\quad n_R =\frac{2\td{h}_v^0}{\td\Delta_0''
r_H\lambda}(w-m_a\td\Omega_a)\,.\ee
%%%
The left-moving and right-moving temperatures $T_L$ and $T_R$ are
then defined by writing the Boltzmann factor as
%%%
\be e^{-(\omega-m_a\Omega_a)/T_H}=e^{-n_L^a/T_L^a- n_R/T_R}\,.\ee
%%%
As a result,
%%%
\be T_L^a=\frac{T_H}{\td\Omega_a-\Omega_a}\,,\quad T_R =
\frac{2\td{h}_v^0}{\td\Delta_0''r_H\lambda}T_H\,.\ee
%%%
In a black hole solution, there should always be a parameter
corresponding to each global charge that the solution may have.
For a rotation $\Omega_a$, the corresponding global charge is
angular momentum, and let's suppose the corresponding parameter in
the solution is given by $\ell_a$. To obtain the extremal limit
for the temperatures, one can take $\ell_a$ to its extremal value
$\td\ell_a$. On the horizon,
%%%
\be\Delta(r_H)=0\quad\Longrightarrow\quad 0=\frac{d\Delta(r_H)}{d
\ell_a}=\frac{\pd\Delta(r_H)}{\pd\ell_a} +\frac{\pd\Delta(r_H)}{
\pd r_H}\frac{d r_H}{d\ell_a}\,.\ee
%%%
Because $\pd\Delta(r_H)/\pd\ell_a$ is finite\footnote{Note
$\pd\Delta(r_H)/\pd\ell_a=0$ corresponds to the case where
$\Delta(r)$ does not contain the parameter $\ell_a$, which in turn
means that $r_H$ is independent of $\ell_a$. This is unlikely to
happen.}, one has in the extremal limit
%%%
\be\frac{\pd\Delta(r_H)}{\pd r_H}\quad\longrightarrow\quad0
\quad\Longrightarrow\quad \frac{d r_H}{d\ell_a} =-\frac{\pd
\Delta(r_H)}{\pd \ell_a}\Big/\frac{\pd\Delta(r_H)}{ \pd r_H}
\quad\longrightarrow\quad\infty\,.\ee
%%%
So in the extremal limit, $T_R= 0$ and
%%%
\bea T_L^a&=&\frac{T_H}{\td\Omega_a-\Omega_a}\Big|_{\ell_a
\rightarrow\td\ell_a}=-\Big(\frac{dT_H}{d\ell_a}\Big/\frac{d
\Omega_a}{d \ell_a} \Big)\Big|_{\ell_a\rightarrow\td\ell_a}\nn\\
&=&-\Big(\frac{\pd T_H}{\pd\ell_a}+\frac{\pd T_H}{\pd r_H} \frac{d
r_H}{d\ell_a}\Big)\Big/\Big(\frac{\pd\Omega_a}{\pd\ell_a}
+\frac{\pd\Omega_a}{\pd r_H}\frac{d r_H}{d\ell_a}\Big)
\Big|_{\ell_a\rightarrow\td\ell_a}\nn\\
&=&-\Big(\frac{\pd T_H}{\pd r_H}\Big/\frac{\pd\Omega_a}{\pd
r_H}\Big) \Big|_{\ell_a\rightarrow\td\ell_a}
=-\frac{\td{T}_H'(r_H)}{\td\Omega_a'} =-\frac{\td
\Delta_0''}{4\pi\td\Omega_a'\td{h}_v^0}\nn\\
&=&\frac{1}{2\pi k^a}\,,\label{cft.def.TL}\eea
%%%
where we have used (\ref{cft.def.ka}). The result
(\ref{cft.def.TL}) was first speculated to be true for general
extremal black holes in four dimensions in \cite{hmns08}. It was
then generalized to solutions in arbitrary dimensions in
\cite{chow.cvetic.lu.pope08} based on all the examples that are
studied. Here we have shown that (\ref{cft.def.TL}) is true for
all known extremal stationary and axisymmetric black holes.

Now by using (\ref{def.ca.general}), (\ref{cft.def.TL}) and
Cardy's formula for the entropy of a unitary conformal field
theory at temperature $T_L$, we find that the microscopic entropy
is given by (no summation over $a$)
%%%
\be S=\frac13 \pi^2\,  c_L^a\, T_L^a=\frac{\cA_{rea}}{4}\,,
\label{cft.def.S}\ee
%%%
where we have identified $c_L^a$ with $c^a$. We see that this
result exactly matches with the Bekenstein-Hawking entropy.

Since the central charge $c^a$ in (\ref{def.ca.general}) only
contains the contribution from the gravitational field, the fact
that (\ref{cft.def.S}) matches with the Bekenstein-Hawking entropy
implies that the non-gravitational contributions to the central
charge(s) are zero. This is consistent with the results found in
\cite{hmns08,ghezelbash09,cmn09}.

\section{Summary}\label{sec.summary}

In this paper, we have calculated the microscopic entropy for all
known extremal stationary and axisymmetric black holes by using
the Kerr/CFT correspondence.

We started by presenting two ansatzs (\ref{metric.general.a1}) and
(\ref{metric.general.a2}) that are general enough to cover all
known stationary and axisymmetric black holes. Then more
constraints on the metrics are introduced from the fact that the
black hole horizons are regular. A common form of the near-horizon
metric (\ref{metric.general.nhem2}) can be derived when the
near-horizon limit is taken for extremal black holes. By using
this near-horizon metric, we explicitly show that the microscopic
entropy calculated by using Cardy's formula exactly matches with
the Bekenstein-Hawking entropy. In this way, we have shown that
the Kerr/CFT correspondence is applicable to all known extremal
stationary and axisymmetric black holes.

For practical reasons, we have only considered the contribution
from the Einstein-Hilbert action to the central charges. And the
match of the microscopic and the macroscopic entropies indicates
that the non-gravitational fields do not contribute to the central
charge(s). Although one can certainly try to repeat the same
process for more complicated theories, such as what has been done
in \cite{acott09}, the calculation will be much more complicated.

Finally, being able to calculate the entropy for a large class of
black holes by using a general method is an encouraging progress.
We hope that the result obtained in this work can help lead to
some true understanding of the microscopic origin of the black
hole entropy.

\section*{Acknowledgement}

I would like to thank Prof. C. N. Pope and Prof. H. L\"u for
helpful discussions, especially for the discussion over the
generality of the ansatzs (\ref{metric.general.a1}) and
(\ref{metric.general.a2}). I also thank the anonymous referee for
his questions and an important reference.

\appendix

\section{Calculating the Central Term $K[\xi,\zeta]$}\label{app.central}

The central term $K[\xi_m^a,\xi_n^a]$ for
(\ref{metric.general.nhem2}) can be calculated by using
(\ref{cov.def.central.minus}) and (\ref{cov.def.kxi.GR}), which
are derived by using the Einstein-Hilbert action alone.

Lets first write down the non-vanishing metric elements in
(\ref{metric.general.nhem2}),\footnote{In this section, we shall
use the capital letter $G$ to denote the full metric
(\ref{metric.general.nhem2}), in order to distinguish it from the
elements $g_{ij}^0$ and $g_{ab}^0$.}
%%%
\bea G_{tt}&=&-A(1+r^2) +k^2 r^2\,,\nn\\
G_{at}&=&G_{t a}=k_a r\,,\nn\\
G_{ab}&=&g_{ab}^0\,,\nn\\
G_{ij}&=&g_{ij}^0\,,\nn\\
G_{rr}&=&\frac{A}{1+r^2}\,,\eea
%%%
where $k_a=g_{ab}^0k^b$, $k^2=g_{ab}^0k^ak^b$ and
$A=2f_r^0/\Delta_0''$. Note $f_r^0=f_r(r_H,\theta^i)$,
$g_{ij}^0=g_{ij}(r_H,\theta^i)$ and
$g_{ab}^0=g_{ab}(r_H,\theta^i)$ are functions of $\theta^i$'s
only, while $\Delta_0''=\Delta''(r_H)$ and $k^a$'s are constant.
Let $(g^{0ab})$ be the inverse of $(g_{ab}^0)$, and $(g^{0ij})$ be
the inverse of $(g_{ij}^0)$, one has
%%%
\bea G^{tt}&=&-\frac1{A(1+r^2)}\,,\nn\\
G^{at}&=&G^{ta}=\frac{k^ar}{A(1+r^2)}\,,\nn\\
G^{ab}&=&g^{0ab}-\frac{k^ak^br^2}{A(1+r^2)}\,,\nn\\
G^{ij}&=&g^{0ij}\,,\nn\\
G^{rr}&=&\frac{1+r^2}{A}\,.\eea
%%%
For later convenience, note that
%%%
\bea\Gamma^t_{ra}&=&-\frac1{2A(1+r^2)}k_a\,,\nn\\
\Gamma^t_{rt}&=&\frac{r}{1+r^2}-\frac{k^2r}{2A(1+r^2)}\,,\nn\\
\Gamma^r_{rr}&=&-\frac{r}{1+r^2}\,,\nn\\
\Gamma^a_{rb}&=&\frac{r}{2A(1+r^2)}k^ak_b\,,\nn\\
\Gamma^i_{rj}&=&0\,,\nn\\
\Gamma^t_{rr}&=&0\,,\nn\\
\Gamma^a_{rt}&=&\frac{1-r^2}{2(1+r^2)}k^a +\frac{k^2r^2}{2A
(1+r^2)}k^a\,.\eea
%%%
Given a particular azimuthal angle $\phi^{\bar{a}}$, and the
Killing vector
%%%
\be \xi_n=-e^{-in\phi^{\bar a}}\,\pd_{\phi^{\bar a}}
-inre^{-in\phi^{\bar a}}\pd_r\,,\ee
%%%
the nontrivial elements of
%%%
\be h_{\mu\nu}(\xi_n)=\cL_{\xi_n} G_{\mu\nu}=\xi_n^\rho \pd_\rho
G_{\mu\nu} +G_{\mu\rho}\pd_\nu\xi_n^\rho +G_{\rho\nu} \pd_\mu
\xi_n^\rho\ee
%%%
are given by
%%%
\bea h_{rr}&=&\xi_n^r \pd_r G_{rr} +2G_{rr}\pd_r\xi_n^r
=-\frac{2ine^{-in\phi^{\bar a}}A}{(1+r^2)^2}\,,\nn\\
h_{ra}&=&G_{rr}\pd_a\xi_n^r=-\frac{n^2re^{-in\phi^{\bar
a}}A}{1+r^2}\delta_{a\bar{a}}\,,\nn\\
h_{tt}&=&\xi_n^r\pd_rG_{tt}=2inr^2 e^{-in\phi^{\bar a}}(A-k^2)\,,\nn\\
h_{ta}&=&\xi_n^r\pd_rG_{t a}+G_{t b}\pd_a\xi_n^b =-in r
e^{-in\phi^{\bar a}}(k_a-k_{\bar a}\delta_{\bar{a}a})\,,\nn\\
h_{ab}&=&G_{ac}\pd_b\xi_n^c+G_{cb}\pd_a\xi_n^c =in e^{-in
\phi^{\bar a}}(g_{a\bar{a}}^0\delta_{\bar{a}b} +g_{b\bar{a}}^0
\delta_{\bar{a}a})\,.\eea
%%%
As a result, $h=0$ and
%%%
\bea h^{rr}&=&G^{rr}G^{rr}h_{rr} =-\frac{2ine^{-in\phi^{\bar a}}}{A}\,,\nn\\
h^{ra}&=&G^{rr}G^{ab}h_{rb}=-n^2re^{-in\phi^{\bar a}}\Big(g^{0a
\bar{a}}-\frac{r^2k^ak^{\bar a}}{A(1+r^2)}\Big)\,,\nn\\
h^{rt}&=&G^{rr}G^{ta}h_{ra}=-\frac{n^2r^2e^{-in\phi^{\bar
a}}}{A(1+r^2)}k^{\bar a}\,,\nn\\
h^{tt}&=&G^{tt}G^{tt}h_{tt} +2G^{tt}G^{ta}h_{ta} +G^{ta} G^{tb}
h_{ab}=\frac{2inr^2e^{-in\phi^{\bar a}}}{A(1+r^2)^2}\,,\nn\\
h^{ta}&=&G^{tt}G^{at}h_{tt} +(G^{tt}G^{ab}+G^{tb}G^{at})h_{tb}
+G^{tb}G^{ac}h_{bc}\nn\\
&=&\frac{inre^{-in\phi^{\bar a}}}{A(1+r^2)}\Big(\frac{1
-r^2}{1+r^2}k^a+k^{\bar a}\delta^{\bar{a}a}\Big)\,,\nn\\
h^{ab}&=&G^{at}G^{bt}h_{tt} +(G^{at}G^{bc}+G^{ac}G^{bt})h_{tc}
+G^{ac}G^{bd} h_{cd}\nn\\
&=&ine^{-in\phi^{\bar a}}\Big[\delta^{a\bar a}g^{0b\bar a}
+\delta^{b\bar a}g^{0a\bar a}-\frac{2r^2k^ak^b}{A(1+r^2)^2}\nn\\
&&\qquad\qquad-\frac{r^2k^{\bar a}(\delta^{a\bar a}k^b
+\delta^{b\bar a}k^a)}{A(1+r^2)}\Big]\,.\eea
%%%
From (\ref{cov.def.kxi.GR}), one has
%%%
\bea k^{rt}&=& \xi_m^t\nabla^r h - \xi_m^t \nabla_\rho h^{r\rho}
+\frac{h}{2}\nabla^t\xi_m^r -h^{t\rho}\nabla_\rho\xi_m^r
+\xi_{m\rho}\nabla^t h^{r\rho}\nn\\
&&-\xi_m^r\nabla^t h+ \xi_m^r \nabla_\rho h^{t\rho} -\frac{h}{2}
\nabla^r\xi_m^t+h^{r\rho}\nabla_\rho\xi_m^t-\xi_{m \rho}\nabla^r
h^{t\rho}\,.\label{cov.def.kxi.GR.value1}\eea
%%%
We are only interested in terms that will lead to $m^3$ when
$m+n=0$ is applied,
%%%
\bea\xi_m^r\nabla_\rho h^{t\rho}&=&\xi_m^r(\pd_\rho h^{t\rho}
+\Gamma^t_{\rho\sigma}h^{\sigma\rho} +\Gamma^\rho_{\rho\sigma}
h^{t\sigma})\nn\\
&\approx&\xi_m^r(\pd_{\bar a}h^{t\bar a}+\pd_r h^{tr}
+2\Gamma^t_{ra}h^{ra}+2\Gamma^t_{rt}h^{rt}
+\Gamma^\rho_{\rho r} h^{tr})\,,\nn\\
&=&\frac{imn^2r^2e^{-i(m+n)\phi^{\bar a}}}{2A(1+r^2)}
\Big(\frac{2r^2-2}{1+r^2}\Big)k^{\bar a}\,,\nn\\
-h^{t\rho}\nabla_\rho\xi_m^r&=&-h^{t\rho}(\pd_\rho\xi_m^r
+\Gamma^r_{\rho\sigma}\xi_m^\sigma)\nn\\
&\approx&-h^{t\bar a}\pd_{\bar a}\xi_m^r -h^{tr}(\pd_r\xi_m^r
+\Gamma^r_{rr}\xi_m^r)\nn\\
&=&\frac{imn^2r^2e^{-i(m+n)\phi^{\bar a}}}{2A(1+r^2)}
\Big(\frac{4m/n-2}{1+r^2}\Big)k^{\bar a}\,,\nn\\
h^{r\rho}\nabla_\rho\xi_m^t&=&h^{r\rho}(\pd_\rho\xi_m^t
+\Gamma^t_{\rho \sigma}\xi_m^\sigma)\nn\\
&\approx&(h^{ra}\Gamma^t_{ar}+h^{rt}\Gamma^t_{tr})\xi_m^r\nn\\
&=&\frac{imn^2r^2e^{-i(m+n)\phi^{\bar a}}}{2A(1+r^2)}
\Big(\frac{r^2-1}{1+r^2}\Big)k^{\bar a}\,,\nn\\
\xi_{m\rho}\nabla^t h^{r\rho}&=&\xi_m^\rho G^{rr}(G^{tt}\nabla_t
h_{r\rho} +G^{ta}\nabla_a h_{r\rho})\nn\\
&=&\xi_m^\rho G^{rr}G^{tt}(\pd_th_{r\rho} -\Gamma^\sigma_{tr}
h_{\sigma\rho} -\Gamma^\sigma_{t\rho}h_{r\sigma})\nn\\
&&+\xi_m^\rho G^{rr}G^{ta}(\pd_ah_{r\rho} -\Gamma^\sigma_{ar}
h_{\sigma\rho} -\Gamma^\sigma_{a\rho}h_{r\sigma})\nn\\
&\approx&\xi_m^r G^{rr}G^{tt}(-\Gamma^{\bar a}_{tr} h_{\bar a r}
 -\Gamma^{\bar a}_{tr}h_{r\bar a})\nn\\
&&+\xi_m^{\bar a}G^{rr}G^{t\bar a}\pd_{\bar a}h_{r\bar a} +\xi_m^r
G^{rr}G^{t\bar a}\pd_{\bar a}h_{rr}\nn\\
&&+\xi_m^r G^{rr}G^{ta}(-\Gamma^{\bar a}_{ar} h_{\bar a r}
-\Gamma^{\bar a}_{ar}h_{r\bar a})\nn\\
&=&\frac{imn^2r^2e^{-i(m+n)\phi^{\bar a}}}{2A(1+r^2)} \Big(
\frac{6-2r^2}{1+r^2}-\frac{2n}{m}\Big)k^{\bar a}\,,\nn\\
-\xi_{m\rho}\nabla^rh^{t\rho}&=&-\xi_{m\rho}G^{rr}
(\pd_rh^{t\rho}+\Gamma^t_{r\sigma}h^{\sigma\rho}
+\Gamma^\rho_{r\sigma}h^{t \sigma})\nn\\
&\approx&-\xi_{mr}G^{rr} (\pd_rh^{tr}+\Gamma^t_{rt}h^{tr}
+\Gamma^t_{r\bar a}h^{\bar a r} +\Gamma^r_{rr}h^{tr})\nn\\
&=&\frac{imn^2r^2e^{-i(m+n)\phi^{\bar a}}}{2A(1+r^2)}
\Big(1-\frac{4}{1+r^2}\Big)k^{\bar a}\,,
\label{cov.def.kxi.GR.value.terms}\eea
%%%
where ``$\approx$" means only terms contributing to $m^3$ are
preserved. The integral in (\ref{cov.def.central.minus}) is done
at $r\rightarrow+\infty$. In this limit, we have from
(\ref{cov.def.kxi.GR.value1}) and
(\ref{cov.def.kxi.GR.value.terms}),
%%%
\be k^{rt}=\frac{i(m-n)n^2e^{-i(m+n)\phi^{\bar a}}}{A}k^{\bar a}
\,.\label{cov.def.kxi.GR.value2}\ee
%%%
Now using (\ref{cov.def.central.minus}) and
(\ref{cov.def.kxi.GR}), and noticing that
%%%
\be\oint(d^{d-2}x)_{\mu\nu}k^{\mu\nu}=\oint2(d^{d-2}x)_{rt}k^{rt}\,,\quad
(d^{d-2}x)_{rt}=\frac12A\sqrt{|g_{ij}^0|}\sqrt{|g_{ab}^0|} \prod_i
d\theta^i \prod_a d\phi^a\,,\ee
%%%
one has
%%%
\bea K[\xi_m^{\bar a},\xi_n^{\bar a}]&=&-\frac{i(m-n)n^2k^{\bar
a}}{16\pi}\oint \sqrt{|g_{ij}^0|}\sqrt{|g_{ab}^0|} \prod_i
d\theta^i \prod_a d\phi^a e^{-i(m+n)\phi^{\bar a}}\nn\\
&=&-\frac{i(m-n)n^2k^{\bar a}}{16\pi}\delta_{m+n}\cA_{rea}\,.
\label{def.value.Kmn}\eea
%%%
Note $\cA_{rea}=\oint \sqrt{|g_{ij}^0|} \sqrt{|g_{ab}^0|} \prod_i
d\theta^i \prod_a d\phi^a$ is the horizon area for both
(\ref{metric.general.c}) and (\ref{metric.general.d}).

\section{The Asymptotic Symmetry Group}\label{app.symmetry}

Asymptotic symmetries are transformations that leave the metric
invariant up to what is allowed by given boundary conditions. One
convenient way to treat asymptotic symmetries is the covariant
phase space method as in \cite{iyer.wald94,iyer.wald95}, which is
also good for exact symmetries. The formalism was first used to
calculate the central charge of conformal symmetries related to a
black hole horizon in \cite{carlip99}. After that, there have been
a lot of further developments. Some examples can be found in
\cite{park01,silva02,barnich.brandt01,barnich.compere07}.

To motivate for the covariant phase space method, one starts with
the classical mechanics (see, e.g.\cite{wald09}). The Lagrangian
is given by $L=L(q, \dtq)$, where $q=q(t)$ describes the classical
trajectory of a particle. For a small variation of the path,
%%%
\be\delta L=\Big(\frac{\pd L}{\pd q} -\frac{d}{dt}\frac{\pd
L}{\pd\dtq}\Big)\delta q +\frac{d}{dt}\Big(\frac{\pd L}{\pd
\dtq}\delta q\Big)\,.\label{cov.vari.L.particle}\ee
%%%
The equation of motion is given by
%%%
\be E=\frac{\pd L}{\pd q} -\frac{d}{dt}\frac{\pd L}{\pd\dtq}
=0\,.\label{cov.def.E}\ee
%%%
When this is linearized, one has
%%%
\be\delta E=\frac{\pd^2L}{\pd q^2}\delta q +\frac{\pd^2L}{\pd\dtq
\pd q}\delta\dtq -\delta\dot{p}=0 \,,\quad p=\frac{\pd L}{\pd\dtq}
\,.\label{cov.linear.E}\ee
%%%
From the boundary term in (\ref{cov.vari.L.particle}), one can
define $\Theta(q,\delta)=p\delta q$ and
%%%
\bea\Omega(q;\delta_1,\delta_2)&=&\delta_1\Theta(q,\delta_2)
-\delta_2\Theta(q,\delta_1)\nn\\
&=&\delta_1p\delta_2q-\delta_2p\delta_1q\,,
\label{cov.def.Omega1}\eea
%%%
where $\delta_1$ and $\delta_2$ stands for two independent
variations. Notice that $\Omega(q;\delta_1,\delta_2)$ is time
independent if both $\delta_1q$ and $\delta_2q$ satisfy
(\ref{cov.linear.E}),
%%%
\be\frac{d\Omega(q;\delta_1,\delta_2)}{dt}
=\delta_1\dtp\delta_2q+\delta_1p\delta_2\dtq
-\delta_2\dtp\delta_1q-\delta_2p\delta_1\dtq =0\,.\ee
%%%
The Hamiltonian of the system can now be defined as
%%%
\be\delta H=\Omega\Big(q;\delta,\frac{d}{dt}\Big)
=\delta\Theta\Big(q,\frac{d}{dt}\Big) -\frac{d}{dt}
\Theta(q,\delta) =\delta p\dtq -\dtp\delta q \,.
\label{cov.def.vari.H}\ee
%%%
Here we have taken the liberty to generalize $\delta$ to other
possible operators, such as $d/dt$. In the case of a curved
spacetime, one might also use the Lie derivative $\cL_\xi$. It
follows that
%%%
\be\dtq=\frac{\pd H}{\pd p}\,,\quad \dtp=-\frac{\pd H}{\pd
q}\,.\ee
%%%
Using generalized coordinates, $\phi^a=\{q,p\}\,,\,a=1,2$, one can
write
%%%
\be\Omega(\phi^a;\delta_1,\delta_2)=\Omega_{ab}\delta_1\phi^a
\delta_2\phi^b\,,\quad (\Omega_{ab})=\left(\begin{matrix}&-1\cr
1&\end{matrix}\right)\,.\label{cov.def.Omega.ab}\ee
%%%
Let $(\Omega^{ab})$ be the inverse of $(\Omega_{ab})$,
%%%
\be(\Omega^{ab})=\left(\begin{matrix}&1\cr-1&\end{matrix}\right)\,,\ee
%%%
the Poisson bracket of any two functions is then given by
%%%
\be\Big\{f\,,\,g\Big\}_{P.B.}=\Omega^{ab}\pd_a f\pd_b g =\frac{\pd
f}{\pd q} \frac{\pd g}{\pd p} -\frac{\pd f}{\pd p}\frac{\pd g}{\pd
q}\,.\ee
%%%
A special example is that, for $f=f(q,p)$,
%%%
\be\frac{d f}{dt}=\frac{\pd f}{\pd q}\dtq +\frac{\pd f}{\pd p}\dtp
=\frac{\pd f}{\pd q} \frac{\pd H}{\pd p} -\frac{\pd f}{\pd
p}\frac{\pd H}{\pd q}=\Big\{f\,,\,H\Big\}_{P.B.}\,.\ee
%%%

For a more general system, there can be more coordinates than just
$\{q,p\}$ and $\Omega_{ab}$ can be more complicated than in
(\ref{cov.def.Omega.ab}). By analogy to (\ref{cov.def.vari.H}),
one can try to construct a charge $Q_\xi$ corresponding to any
symmetric transformation $\delta_\xi$,
%%%
\be\delta Q_\xi=\Omega(\phi^a;\delta,\delta_\xi)=\Omega_{ab}
\delta\phi^a \delta_\xi\phi^b\,.\label{cov.def.Q.mechanics}\ee
%%%
To make $Q_\xi$ a physically meaningful charge, the variation
(\ref{cov.def.Q.mechanics}) needs to be integrable and
$\Omega(\phi^a;\delta,\delta_\xi)$ needs to be constant in time.
This will put extra constraints on $\delta\phi^a$ and
$\delta_\xi\phi^a$, just as in the case above. Given two charges
as defined in (\ref{cov.def.Q.mechanics}), the Poisson bracket is
%%%
\be\Big\{Q_\xi\,,\,Q_\zeta\Big\}_{P.B.}=\Omega^{ab} \frac{\delta
Q_\xi}{\delta\phi^a} \frac{\delta Q_\zeta}{\delta\phi^b}
=\Omega(\phi^a;\delta_\zeta,\delta_\xi)\,.
\label{cov.def.PB.mechanics}\ee
%%%
This result will play a central role in the treatment that
follows.

Now consider a system with the Lagrangian density $\cL=\cL(\phi^a,
\pd_\mu\phi^a, \pd_\mu\pd_\nu\phi^a,\cdots)$. The actions is
%%%
\be S=\int_\cM\bL\,,\quad \bL =\cL\sqrt{|g|}\;d^nx
=\cL\ast\one\,.\label{action.general}\ee
%%%
A symmetric transformation should leave the integrand $\bL$
invariant or up to a total derivative which integrates to zero,
%%%
\be\delta_\epsilon\bL=d\bM_\epsilon\,,\quad \delta S=\int_\cM
d\bM_\epsilon=\oint_{\pd\cM} \bM_\epsilon=0\,. \label{def.sym}\ee
%%%
On the other hand,
%%%
\be\delta_\epsilon\bL=E_a\delta_\epsilon\phi^a\ast\one +d{\bf
\Theta}(\phi^a,\delta_\epsilon)\,,\label{vari.L} \ee
%%%
where all the terms involving a derivative on
$\delta_\epsilon\phi^a$ have been moved into the $d\bf\Theta$
term. It is easy to see that $E_a=0$ is the usual Euler-Lagrange
equation for $\phi^a$. From (\ref{def.sym}) and (\ref{vari.L}),
one can define a Noether current,
%%%
\be\bJ_\epsilon={\bf\Theta}(\phi^a,\delta_\epsilon)
-\bM_\epsilon\,,\label{def.J.noether}\ee
%%%
which becomes a closed form when the equations of motion are
satisfied, $d\bJ_\epsilon =-E_a\cdot \delta_\epsilon \phi^a \ast
\one$. So when $E_a=0$, one should locally have
$\bJ_\epsilon=d\bQ_\epsilon$, with $\bQ_\epsilon$ being some $n-2$
form. Now with appropriate boundary conditions, a conserved charge
can be defined as
%%%
\be Q_\epsilon=\int_V d\bQ_\epsilon =\oint_{\pd V}\bQ_\epsilon
\,,\label{def.Q.noether}\ee
%%%
where $V$ is a space-like slice of the spacetime manifold $\cM$.
The charge $\bQ_\epsilon$ is defined up to an arbitrary closed
form, but this ambiguity drops out in (\ref{def.Q.noether}).

For a transformation generated by the Lie derivative,
$\delta_\xi\phi^a =\cL_\xi\phi^a$, one has
%%%
\bea\delta_\xi\bL&=&E_a\cdot\cL_\xi\phi^a\ast\one +d{\bf\Theta}
(\phi^a,\cL_\xi)\nn\\
&=&\cL_\xi\bL=d(i_\xi\bL)\,.\label{cov.trans.L}\eea
%%%
The Noether current (\ref{def.J.noether}) is
%%%
\be\bJ_\xi={\bf\Theta}(\phi^a,\cL_\xi)-i_\xi\bL\,.
\label{cov.def.J.noether}\ee
%%%
By analogy to (\ref{cov.def.Omega1}), one can define
%%%
\bea\Omega(\phi^a;\delta_1,\delta_2)&=&\int_V \bw(\phi^a;
\delta_1,\delta_2)\,,\label{cov.def.Omega2}\\
\bw(\phi^a;\delta_1,\delta_2)&=&\delta_1 {\bf \Theta}(\phi^a,
\delta_2) -\delta_2{\bf \Theta}(\phi^a, \delta_1)\,.
\label{cov.def.w}\eea
%%%
The quantity $\Omega(\phi^a;\delta_1,\delta_2)$ is conserved if
%%%
\be d\bw(\phi^a;\delta_1,\delta_2)=0\quad\Longrightarrow\quad
\oint_{\pd\cM}\bw=\int_\cM d\bw=0\,.\label{cov.constraints}\ee
%%%
Notice that,
%%%
\bea0&=&(\delta_1\delta_2-\delta_1\delta_2)(\cL\ast\one)
\quad\Longleftrightarrow\quad \delta_1\delta_2\phi^a
=\delta_1\delta_2\phi^a\,,\label{cov.asp.delta12}\\
&=& (\delta_1E_a\delta_2\phi^a -\delta_2E_a\delta_1\phi^a)
\ast\one+d\bw(\phi^a;\delta_1,\delta_2)\,.\eea
%%%
As a result,
%%%
\be d\bw(\phi^a;\delta_1,\delta_2)=0 \quad\Longrightarrow\quad
\delta_1E_a =\delta_2 E_a=0\,.\ee
%%%
So $\delta_1 \phi^a$ and $\delta_2\phi^a$ must both satisfy the
linearized equations of motion for $\phi^a$, in order that $\Omega
(\phi^a;\delta_1,\delta_2)$ can be constant in time. When this
condition is satisfied, one can try to construct a charge
corresponding to $\delta_\xi=\cL_\xi$, by analogy to
(\ref{cov.def.vari.H}),
%%%
\be\delta Q_\xi=\Omega(\phi^a;\delta,\cL_\xi)=\int_V
\bw(\phi^a;\delta,\cL_\xi)\,. \label{cov.def.vari.Qxi.1}\ee
%%%
The variation of the Noether current (\ref{cov.def.J.noether}) is
%%%
\bea\delta\bJ_\xi&=&\delta{\bf\Theta}(\phi^a,\cL_\xi)
-i_\xi\delta\bL\nn\\
&=&\delta{\bf\Theta}(\phi^a,\cL_\xi) -\cL_\xi{\bf\Theta}
(\phi^a,\delta) +d\Big[i_\xi{\bf\Theta}(\phi^a,\delta)\Big]\,,
\label{cov.vari.neother}\eea
%%%
where the second line is obtained for $E_a=0$. As a result,
%%%
\bea\bw(\phi^a;\delta,\cL_\xi)&=&\delta{\bf\Theta}(\phi^a,\cL_\xi)
-\cL_\xi{\bf\Theta}(\phi^a, \delta)=d\bk_\xi(\phi^a,\delta)\,,\nn\\
\Longrightarrow\quad\delta Q_\xi&=&\oint_{\pd V} \bk_\xi (\phi^a,
\delta)\,,\label{cov.def.vari.Qxi.2}\eea
%%%
with
%%%
\be\bk_\xi(\phi^a,\delta)=\delta\bQ_\xi-i_\xi{\bf\Theta} (\phi^a,
\delta)\,.\label{cov.def.kxi}\ee
%%%
Note that $\delta(\cL_\xi \phi^a) =\cL_\xi(\delta\phi^a)$, so both
$\delta$ and $\cL_\xi$ satisfy the assumption made about the
operators $\delta_1$ and $\delta_2$ in (\ref{cov.asp.delta12}).
From (\ref{cov.def.vari.Qxi.2}),
%%%
\be Q_\xi(\phi)=\int_{\bar\phi}^\phi\delta Q_\xi +Q_\xi(\bar
\phi)=\int_{\bar \phi}^\phi \oint_{\pd V} \bk_\xi(\phi^a,\delta)
+Q_\xi(\bar \phi)\,,\label{cov.def.Qxi}\ee
%%%
where $Q_\xi(\bar\phi)$ is the value of the charge on a given
background. For the charge $Q_\xi(\phi)$ to be well defined, one
expects the integral to be finite. Now given two such charges (say
$Q_\xi$ and $Q_\zeta$), the Poisson bracket is found by analogy to
(\ref{cov.def.PB.mechanics}),
%%%
\bea\Big\{Q_\xi\,,\,Q_\zeta\Big\}_{P.B.}&=&\Omega(\phi^a;\cL_\zeta,
\cL_\xi)=\oint_{\pd V} \bk_\xi(\phi^a,\cL_\zeta)\,.
\label{cov.def.PB.GR}\eea
%%%
It was shown in \cite{brown.henneaux86a,brown.henneaux86b} that
with appropriate boundary conditions, the Poisson bracket
$\{Q_\xi\,, \,Q_\zeta\}_{P.B.}$ of any differentiable generators
$Q_\xi$ and $Q_\zeta$ takes the form
%%%
\be\Big\{Q_\xi\,,\,Q_\zeta\Big\}_{P.B.}=Q_{[\xi,\zeta]} +K[\xi,
\zeta]\,,\label{cov.def.PB.general}\ee
%%%
where $K[\xi,\zeta]$ is a potential central extension to the
algebra. It is demonstrated in \cite{brown.henneaux86b} that a
constant shift in the charges will not affect the nontrivial part
of $K[\xi,\zeta]$. Using this, we can shift the charges by some
constant and let $Q_{[\xi,\zeta]}(\bar\phi)=0$ in a chosen
background. Then we get
%%%
\be K[\xi, \zeta]=\Big\{Q_\xi\,,\,Q_\zeta\Big\}_{P.B.} =\oint_{\pd
V} \bk_\xi(\bar\phi^a,\cL_\zeta)\,. \label{cov.def.central}\ee
%%%
Note that if instead of using (\ref{cov.def.vari.Qxi.1}), had we
chosen to define
%%%
\be\delta Q_\xi=-\Omega(\phi^a;\delta,\cL_\xi)=-\int_V
\bw(\phi^a;\delta,\cL_\xi)\,,\label{cov.def.vari.Qxi.3}\ee
%%%
we would have got
%%%
\be K[\xi, \zeta]=\Big\{Q_\xi\,,\,Q_\zeta\Big\}_{P.B.} =-\Omega
(\phi^a;\cL_\xi,\cL_\zeta) =-\oint_{\pd V} \bk_\xi
(\phi^a,\cL_\zeta)\,. \label{cov.def.central.minus}\ee
%%%
This result was used in the calculation of the Kerr/CFT
correspondence \cite{ghss08}.

In the case of pure gravity supplemented with a cosmological
constant, the Lagrangian density is given by
%%%
\be\cL=\frac{R-2\Lambda}{16\pi}\,.\ee
%%%
For an infinitesimal variation of the metric,
%%%
\be\delta\bL=\frac{1}{16\pi}\Big(-R^{\mu\nu} +\frac{R-2
\Lambda}{2} g^{\mu\nu} +\nabla^\mu \nabla^\nu -g^{\mu\nu}
\nabla_\rho \nabla^\rho \Big)\delta g_{\mu\nu}\ast\one\,.
\label{vari.L.pureE.Lambda}\ee
%%%
Einstein's equations are
%%%
\bea E^{\mu\nu}&=&R^{\mu\nu} -\frac{R-2\Lambda}{2}
g^{\mu\nu}=0\,,\label{cov.eoms.R.Lambda}\\
\Longrightarrow\quad R_{\mu\nu}&=&\frac{2\Lambda}{n-2}
g_{\mu\nu}\,,\quad R=\frac{2n\Lambda}{n-2}\,.\eea
%%%
When (\ref{cov.eoms.R.Lambda}) is linearized, one has
%%%
\bea0=\delta E_{\mu\nu}&=&\frac12\Big[\nabla^\rho(\nabla_\mu
h_{\nu\rho}+\nabla_\nu h_{\mu\rho})-\pd^\rho\pd_\rho
h_{\mu\nu}-\nabla_\mu\nabla_\nu h\Big]\nn\\
&&-\frac12\Big[\nabla_\mu\nabla_\nu h^{\mu\nu} -\pd^\rho\pd_\rho h
-R^{\rho \sigma} h_{\rho\sigma}\Big]g_{\mu\nu} -\frac{R
-2\Lambda}{2} h_{\mu\nu}\,,\label{cov.eoms.linearized}\eea
%%%
where $h_{\mu\nu}=\delta g_{\mu\nu}$ and $h=g^{\mu\nu}h_{\mu\nu}$.
Taking the trace of (\ref{cov.eoms.linearized}), one has
%%%
\be\nabla_\mu\nabla_\nu h^{\mu\nu} -\pd^\rho\pd_\rho h -R^{\mu\nu}
h_{\mu\nu}=0\,.\ee
%%%
From (\ref{cov.trans.L}),
%%%
\bea{\bf\Theta}(g_{\mu\nu},\delta)&=&\frac{1}{16\pi} (d^{n-1}
x)_\mu\Big[\nabla_\nu h^{\mu\nu} -\nabla^\mu h\Big]\,,\nn\\
\Longrightarrow\quad i_\xi{\bf\Theta}(g_{\mu\nu},\delta)
&=&\frac{1}{16\pi} (d^{n-2}x)_{\mu\nu}2\xi^\nu(\nabla_\nu
h^{\mu\nu}-\nabla^\mu h)\nn\\
&=&\frac{1}{16\pi}(d^{n-2}x)_{\mu\nu}(-I_{\Theta_\xi}^{\mu\nu})\,,
\label{cov.def.ixi.Theta}\eea
%%%
where
%%%
\be I_{\Theta_\xi}^{\mu\nu}=\xi^\mu\nabla_\rho h^{\nu\rho} -
\xi^\nu \nabla_\rho h^{\mu\rho}+\xi^\nu\nabla^\mu
h-\xi^\mu\nabla^\nu h \,.\ee
%%%
The Noether current (\ref{cov.def.J.noether}) is
%%%
\bea\bJ_\xi&=&\frac{1}{16\pi} (d^{n-1} x)_\mu\Big[\nabla^\nu
\nabla^\mu\xi_\nu+\pd^\rho\pd_\rho\xi^\mu-2\nabla^\mu\nabla^\nu\xi_\nu
-(R-2\Lambda)\xi^\mu\Big]\nn\\
&=&-\frac{1}{16\pi} (d^{n-1} x)_\mu\nabla_\nu\Big[\nabla^\mu
\xi^\nu-\nabla^\nu\xi^\mu\Big]\,,\nn\\
\Longrightarrow\quad\bQ_\xi&=&-\frac{1}{16\pi}(d^{n-2}x)_{\mu\nu}
(\nabla^\mu \xi^\nu-\nabla^\nu\xi^\mu)\,,
\label{cov.def.Q.R.Lambda}\eea
%%%
where we have used (\ref{cov.eoms.R.Lambda}). Note that $\delta
\bQ_\xi=\frac{1}{16\pi}(d^{n-2}x)_{\mu\nu}I_{Q_\xi}^{\mu\nu}$,
with
%%%
\bea I_{Q_\xi}^{\mu\nu}=-\frac{h}{2}(\nabla^{\mu}\xi^{\nu}
-\nabla^{ \nu}\xi^{\mu})+h^{\mu\rho}\nabla_\rho\xi^\nu-h^{\nu\rho}
\nabla_\rho\xi^\mu\nn\\
-(\nabla^{\mu}h^{\nu\rho}-\nabla^{\nu}h^{\mu\rho})\xi_\rho\,.
\label{cov.def.delta.Qxi}\eea
%%%
From (\ref{cov.def.kxi}), one gets that
%%%
\bea\bk_\xi(g_{\mu\nu},\delta)&=&\frac{1}{16\pi}
(d^{n-2}x)_{\mu\nu}k^{\mu\nu}\,,\nn\\
k^{\mu\nu}=I_{Q_\xi}^{\mu\nu} +I_{\Theta_\xi}^{\mu\nu}&=&
\xi^\nu\nabla^\mu h - \xi^\nu \nabla_\rho h^{\mu\rho}
+\frac{h}{2}\nabla^{ \nu}\xi^{\mu} -h^{\nu\rho}\nabla_\rho
\xi^\mu+\xi_\rho\nabla^{\nu}h^{\mu\rho}\nn\\
&&-(\mu\leftrightarrow\nu) \,. \label{cov.def.kxi.GR}\eea
%%%
This result matches with that given in \cite{cmn09} up to a
trivial term. Note \cite{ghss08} uses a formula for
$\bk_\xi(g_{\mu\nu},\delta)$ with the opposite sign, for which to
make sense, we need to use (\ref{cov.def.vari.Qxi.3}) and
(\ref{cov.def.central.minus}).

To clarify the notations involved, note that we write a $p$-form
as
%%%
\be\bw_p=\frac{1}{p!}w_{\mu_1\cdots\mu_p}
dx^{\mu_1}\wedge\cdots\wedge dx^{\mu_p} \,.\ee
%%%
Its Hodge-$\ast$ dual is defined by (note $|\epsilon_{\cdots}|
=\sqrt{|g|}$)
%%%
\be\ast \bw_p=w^{\mu_1\cdots\mu_p}\frac{1}{p!(n-p)!}
\epsilon_{\mu_1\cdots\mu_p\nu_1\cdots\nu_{n-p}}
dx^{\nu_1}\wedge\cdots\wedge dx^{\nu_{n-p}} \,.\ee
%%%
One can also write it as
%%%
\bea\ast \bw_p&=&(d^{n-p}x)_{\mu_1\cdots\mu_p} w^{\mu_1
\cdots\mu_p}\,,\label{cgr.def.hodge.new.form}\\
(d^{n-p}x)_{\mu_1\cdots\mu_p}&=&\frac{1}{p!(n-p)!}
\epsilon_{\mu_1\cdots\mu_p\nu_1\cdots\nu_{n-p}}
dx^{\nu_1}\wedge\cdots\wedge dx^{\nu_{n-p}}\,.\eea
%%%
With this, Stokes's theorem $\int_{\Sigma}d\ast
\bw_p=\oint_{\pd\Sigma}\ast \bw_p$ can be written as
%%%
\be\int_{\Sigma}(d^{n-p+1}x)_{\mu_2\cdots\mu_p}\nabla_{\mu_1}
w^{\mu_1\mu_2\cdots\mu_p}=\oint_{\pd\Sigma} (d^{n
-p}x)_{\mu_2\cdots\mu_p\mu_1} w^{\mu_1\mu_2\cdots\mu_p}\,.
\label{cgr.def.stokes.new.form}\ee
%%%

\section{Some Examples}\label{app.examples}

In this section, we use some examples to illustrate some major
points made in the main context. The majority of the examples have
been studied in
\cite{lu.mei.pope08c,chow.cvetic.lu.pope08,lu.mei.pope.vazquez-poritz09}.
Here we discuss them again by using the new perspective that we
have gained from the present work. Since all the calculations
after (\ref{metric.general.c}) and (\ref{metric.general.d}) evolve
in a straight forward manor, our goal here is to show that all the
examples can be put into the form of either
(\ref{metric.general.c}) or (\ref{metric.general.d}) as $\hat
r\rightarrow r_H$.

One intriguing result we find is that a surprisingly large number
of solutions are exactly of the form (\ref{metric.general.c}) with
$h_\cA=h_\chi^a=h_{tt}=0$. This feature could be helpful when one
is trying to look for new solutions.

\subsection{Kerr-NUT-AdS Solutions in Diverse Dimensions}

Lets start with examples studied in \cite{lu.mei.pope08c}.

The first example is the Kerr-AdS solution in four dimensions
\cite{carter68},
%%%
\bea ds^2&=&\rho^2\Big(\frac{d\hat r^2}{\Delta} + \frac{d
\theta^2}{\Delta_{\theta}}\Big)+\frac{\Delta_\theta \sin^2
\theta}{\rho^2}\Big(a d\hat t-\frac{\hat r^2+a^2}{\Xi}d\hat
\phi\Big)^2 - \frac{\Delta}{\rho^2} (d\hat t - \frac{a
\sin^2\theta}{\Xi} d\hat \phi\Big)^2\,,\nn\\
\rho^2&=& \hat r^2 + a^2 \cos^2\theta\,,\qquad
\Delta=(\hat r^2 + a^2) (1 + \hat r^2\ell^{-2}) - 2 M \hat r\,,\nn\\
&& \Delta_\theta = 1 - a^2\ell^{-2} \cos^2\theta\,,\qquad \Xi=1 -
a^2\ell^{-2}\,.\eea
%%%
It is is a solution to the equations of motion $R_{\mu\nu}=-
3\ell^{-2}\, g_{\mu\nu}$. Comparing with (\ref{metric.feature1})
and (\ref{metric.feature.def.A}), it is easy to see that
%%%
\bea\cA&=&d\hat t-\frac{a\sin^2\theta}{\Xi}d\hat\phi
+\frac{\rho^2}{\Delta}dr\nn\\
&=&d\hat t-\frac{a\sin^2\theta}{\Xi}d\hat\phi +\frac{r^2
+a^2-a^2\sin^2\theta}{\Delta}dr\,,\nn\\
\Longrightarrow\quad h_v&=&r^2+a^2\,,\quad h_\phi=a\,\Xi\,,\quad
h_\cA=0\,.\eea
%%%
One sees that the metric is exactly of the form
(\ref{metric.general.c}) with $h_\cA=h_\chi^\phi=h_{tt}=0$.

The second example is the five-dimensional rotating black hole
with $S^3$ horizon topology. The solutions was obtained by
Hawking, Hunter and Taylor-Robinson
\cite{hawking.hunter.taylor98}, satisfying the equations of motion
$R_{\mu\nu}= -4\ell^{-2}\, g_{\mu\nu}$. The metric, which
generalizes the Ricci-flat rotating black hole of Myers and Perry
\cite{myers.perry86}, is given by
%%%
\bea ds^2&=& - \frac{\Delta}{\rho^2} (d\hat t -
\frac{a\sin^2\theta}{\Xi_a} d\phi_1 - \frac{b\cos^2\theta}{\Xi_b}
d\phi_2)^2 + \frac{\Delta_\theta \sin^2\theta}{\rho^2}
(a d\hat t - \frac{(\hat r^2 + a^2)}{\Xi_a} d\phi_1)^2\nn\\
&&+\frac{\Delta_\theta \cos^2\theta}{\rho^2} (b d\hat t -
\frac{(\hat r^2 + b^2)}{\Xi_b} d\phi_2)^2 +
\frac{\rho^2}{\Delta} d\hat r^2 + \frac{\rho^2}{\Delta_\theta} d\theta^2\\
&& +\frac{1 + \hat r^2\ell^{-2}}{\hat r^2\rho^2} \Big(a b d\hat t
- \frac{b(\hat r^2 + a^2) \sin^2\theta}{\Xi_a} d\phi_1 -
\frac{a(\hat r^2 + b^2)\cos^2\theta}{\Xi_b} d\phi_2\Big)^2 \,,\nn
\eea
%%%
where
%%%
\bea &&\Delta = \frac{1}{\hat r^2}(\hat r^2 + a^2)(\hat r^2 + b^2)
(1 + \hat r^2 \ell^{-2}) - 2M\,,\quad \Delta_\theta=1 -
a^2\ell^{-2} \cos^2\theta - b^2\ell^{-2} \sin^2\theta\,,\nn\\
&&\rho^2 = \hat r^2 + a^2\cos^2\theta + b^2\sin^2\theta\,,\qquad
\Xi_a=1-a^2\ell^{-2}\,,\qquad \Xi_b=1 - b^2 \ell^{-2}\,. \eea
%%%
Note that in this coordinate system, the metric is asymptotic to
AdS$_5$ in a rotating frame, with angular velocities
$\Omega_{\phi_1}^\infty=-a \ell^{-2}$ and
$\Omega_{\phi_2}^\infty=-b \ell^{-2}$. By letting
%%%
\be \phi_1\rightarrow\phi_1-a\ell^{-2}\hat t\,,\quad
\phi_2\rightarrow\phi_2-b\ell^{-2}\hat t\,,\ee
%%%
one can change to an asymptotically static coordinate system. The
metric is now given by{\small
%%%
\bea ds^2&=& - \frac{\Delta}{\rho^2}\left[\Big(1+\frac{a^2
\ell^{-2} \sin \theta^2}{\Xi_a} +\frac{b^2\ell^{-2}\cos
\theta^2}{\Xi_b}\Big) d\hat t - \frac{a\sin^2\theta}{\Xi_a}
d\phi_1 - \frac{b\cos^2 \theta}{\Xi_b}d\phi_2\right]^2\nn\\
&&+\frac{\rho^2}{\Delta} d\hat r^2+ \frac{\Delta_\theta \sin^2
\theta(\hat r^2+a^2)^2}{\rho^2 \Xi_a^2} \Big(d\phi_1 -\frac{a(1
+\hat r^2 \ell^{-2})}{\hat r^2+a^2}d\hat t\Big)^2 \nn\\
&&+ \frac{\rho^2}{\Delta_\theta} d\theta^2+\frac{\Delta_\theta
\cos^2\theta(\hat r^2+b^2)^2}{\rho^2 \Xi_b^2} \Big(d\phi_2
-\frac{b(1+\hat r^2\ell^{-2})}{\hat r^2+b^2}d\hat t\Big)^2\nn\\
&& +\frac{a^2b^2(1+\hat r^2\ell^{-2})}{\hat r^2 \rho^2}
\Bigg\{\frac{(\hat r^2+a^2)\sin^2\theta}{a\Xi_a} \Big( d\phi_1
-\frac{a(1 +\hat r^2 \ell^{-2})}{\hat r^2+a^2}d\hat t\Big)\nn\\
&&\qquad\qquad\qquad+\frac{(\hat
r^2+b^2)\cos^2\theta}{b\Xi_b}\Big(d\phi_2 -\frac{b(1+\hat
r^2\ell^{-2})}{\hat r^2+b^2}d\hat t\Big)\Bigg\}^2\,.
\label{cft.eg.5d.kerr.ads.metric}\eea}
%%%
From (\ref{metric.feature1}) and (\ref{metric.feature.def.A}),
%%%
\bea\cA&=&\Big(1+\frac{a^2 \ell^{-2} \sin \theta^2}{\Xi_a}
+\frac{b^2\ell^{-2}\cos \theta^2}{\Xi_b}\Big) d\hat t\nn\\
&&- \frac{a\sin^2\theta}{\Xi_a} d\phi_1 - \frac{b\cos^2
\theta}{\Xi_b}d\phi_2+\frac{\rho^2}{\Delta}dr\,.\eea
%%%
Comparing (\ref{cft.eg.5d.kerr.ads.metric}) with
(\ref{metric.general.c}), we find
%%%
\bea h_v&=&\frac{(\hat r^2+a^2)(\hat r^2+b^2)}{\hat r^2}\,,\quad
h_1=\frac{a(1+\hat r^2\ell^{-2})}{\hat r^2+a^2}h_v\,,\nn\\
h_2&=&\frac{b(1+\hat r^2\ell^{-2})}{\hat r^2+b^2}h_v\,,\quad
h_\cA=0\,.\eea
%%%
It is easy to see that (\ref{cft.eg.5d.kerr.ads.metric}) is of the
form (\ref{metric.general.c}) with $h_\cA=h_\chi^a=h_{tt}=0$.

In the following, we shall consider the general Kerr-NUT-AdS
solutions found in \cite{chen.lu.pope06b}, which solve the
Einstein equation $R_{\mu\nu}= -(d-1)\ell^{-2}\, g_{\mu\nu}$. The
case of Kerr-AdS solutions have been studied in
\cite{lu.mei.pope08c} and \cite{chow.cvetic.lu.pope08}. Since the
NUT parameters will not affect anything in the process, here we
will include them as well. Also, we will choose to write the
metric by analogy to (40) and (48) in \cite{chen.lu.pope06b},
which specialized to seven and six dimensions respectively. In
even dimensions, $d=2n$, the metric is given by
%%%
\bea ds_{2n}^2&=&\sum_{i=1}^n\Big(\frac{f_idx_i^2}{X_i}
+\frac{X_i}{f_i}\cA_i^2\Big)\,,\quad f_i=\prod_{j\neq
i}(x_i^2-x_j^2)\,,\label{cft.eg.even.d.kerr.nut.ads}\\
\cA_i&=&dt+\sum_{j\neq i}x_j^2d\phi_1 +\sum_{j,k\neq i} x_j^2
x_k^2 d\phi_2+\cdots+\prod_{j\neq i}x_j^2d\phi_{n-1}\,,\nn\\
X_i&=&2M_i x_i+\sum_{j=0}^{n-1}c_{2j}^{~}x_i^{2j}+g^2x_i^{2n}
\,.\eea
%%%
In odd dimensions, $d=2n+1$, the metric is given by
%%%
\be ds_{2n+1}^2=ds_{2n}^2+\frac{c_n}{\prod_{i=1}^nx_i^2}
\cA_n^2\,,\label{cft.eg.odd.d.kerr.nut.ads}\ee
%%%
with
%%%
\bea\cA_n&=&dt+\sum_{i=1}^nx_i^2d\phi_1 +\sum_{i,j=1}^nx_i^2 x_j^2
d\phi_2+\cdots+\prod_{i=1}^nx_i^2d\phi_n\,,\nn\\
X_i&=&(-1)^{\frac{d-1}2}\frac{c_n}{x_i^2}+2M_i +\sum_{j=1}^{n-1}
c_{2j}^{~}x_i^{2j}+g^2x_i^{2n}\,,\nn\\
\cA_{i\neq1}&=&dt+\sum_{j\neq 1,i}x_j^2d\phi_1 +\sum_{j,k\neq 1,i}
x_j^2x_k^2 d\phi_2+\cdots+\prod_{j\neq1,i}x_j^2d\phi_{n-2}\nn\\
&&-r^2\Big(d\phi_1 +\sum_{j\neq 1,i} x_j^2 d\phi_2 +\cdots
+\prod_{j\neq 1,i}x_j^2d\phi_{n-1}\Big)\nn\\
&=&dt-r^2d\phi_1+\sum_{j\neq 1,i}x_j^2(d\phi_1-r^2d\phi_2)+\cdots\nn\\
&&+\prod_{j\neq 1,i}x_j^2\Big(d\phi_{n-2}-r^2d\phi_{n-1}\Big)\,.
\label{cft.eg.kerr.nut.ads.Ai}\eea
%%%
Note we have wick rotated the radial direction
$r^2\rightarrow-x_1^2$ so that the metrics
(\ref{cft.eg.even.d.kerr.nut.ads}) and
(\ref{cft.eg.odd.d.kerr.nut.ads}) can be put into a compact form.
To get back to the Lorentzian signature black hole metric, one
needs to wick rotate back, $x_1^2\rightarrow-r^2$. Especially, one
has
%%%
\bea f_1&=&(-1)^{n-1}\td f_1(r)\,,\quad X_1=(-1)^n X(r)\,,\nn\\
\td f_1(r)&=&r^{2(n-1)}+r^{2(n-2)}\sum_{j>1}x_j^2 +r^{2(n-3)}
\sum_{j,k>1}x_j^2 x_k^2+\cdots+\prod_{j>1}x_j^2\,,\nn\\
X(r)&=&g^2r^{2n}+\cdots\,.\eea
%%%
Now from (\ref{metric.feature1}) and (\ref{metric.feature.def.A}),
one has for both (\ref{cft.eg.even.d.kerr.nut.ads}) and
(\ref{cft.eg.odd.d.kerr.nut.ads}),
%%%
\be\cA=\cA_1+\frac{\td f_1dr}{X}\,.\ee
%%%
As a result, for both even and odd dimensions ($i\leq n-1$),
%%%
\be h_v=r^{2(n-1)}\,,\quad h_i=r^{2(n-1-i)}\,,\quad h_\cA=0\,.\ee
%%%
From (\ref{cft.eg.kerr.nut.ads.Ai}),
%%%
\bea\cA_{i\neq1}&=&dt-\frac{h_v}{h_1}d\phi_1+\sum_{j\neq
1,i}x_j^2\Big[\Big(d\phi_1-\frac{h_1}{h_v}dt\Big)
-r^2\Big(d\phi_2 -\frac{h_2}{h_v}dt\Big)\Big]\nn\\
&&+\cdots+\prod_{j\neq 1,i}x_j^2\Big[\Big(d\phi_{n-2}
-\frac{h_{n-2}}{h_v}dt\Big) -r^2\Big(d\phi_{n-1}
-\frac{h_{n-1}}{h_v}dt\Big)\Big]\,.\eea
%%%
In odd dimensions, we also have
%%%
\bea\cA_n&=&dt+\sum_{i=1}^nx_i^2d\phi_1 +\sum_{i,j=1}^nx_i^2 x_j^2
d\phi_2+\cdots+\prod_{i=1}^nx_i^2d\phi_n\,,\nn\\
&=&dt+\sum_{j\neq 1,i}x_j^2d\phi_1 +\sum_{j,k\neq 1,i}
x_j^2x_k^2 d\phi_2+\cdots+\prod_{j\neq1,i}x_j^2d\phi_{n-1}\nn\\
&&-r^2\Big(d\phi_1 +\sum_{j\neq 1,i} x_j^2 d\phi_2 +\cdots
+\prod_{j\neq 1,i}x_j^2d\phi_n\Big)\nn\\
&=&dt-r^2d\phi_1+\sum_{j\neq 1,i}x_j^2(d\phi_1-r^2d\phi_2)+\cdots\nn\\
&&+\prod_{j\neq1,i}x_j^2\Big(d\phi_{n-1}-r^2d\phi_n\Big)\nn\\
&=&dt-\frac{h_v}{h_1}d\phi_1+\sum_{j\neq
1,i}x_j^2\Big[\Big(d\phi_1-\frac{h_1}{h_v}dt\Big)
-r^2\Big(d\phi_2 -\frac{h_2}{h_v}dt\Big)\Big]\nn\\
&&+\cdots+\prod_{j\neq 1,i}x_j^2\Big[\Big(d\phi_{n-1}
-\frac{h_{n-1}}{h_v}dt\Big) -r^2\Big(d\phi_n
-\frac{h_{n-1}}{r^2h_v}dt\Big)\Big]\,.\eea
%%%
So it is obvious that both (\ref{cft.eg.even.d.kerr.nut.ads}) and
(\ref{cft.eg.odd.d.kerr.nut.ads}) can be put into the form of
(\ref{metric.general.c}), with $h_\cA=h_\chi^a=h_{tt}=0$.

\subsection{Extremal Static Black Holes in Supergravity Theories}

Here we turn to the examples studied in
\cite{lu.mei.pope.vazquez-poritz09}. A key feature here is that
all the solutions are charged but static. In order to use the
Kerr/CFT correspondence, which only works with rotating black
holes, the strategy used in \cite{lu.mei.pope.vazquez-poritz09} is
to lift the charged static solutions into higher dimensions by
using some consistent Kaluza-Klein reduction procedure. The
electric charges of the static black holes then acquire the
interpretation of rotations in the internal dimensions after the
lifting.

Here we will discuss the same examples from the perspective of
using (\ref{metric.general.c}), but we will still be using the
same strategy as employed in \cite{lu.mei.pope.vazquez-poritz09}.
For this purpose, we start with the various reduction ansatz given
in \cite{10author}:
%%%
\begin{itemize}
\item For the $S^5$ reduction of type IIB supergravity, the ansatz
for the ten-dimensional metric is
%%%
\be ds_{10}^2 = \sqrt{\wtd\Delta}\, ds_5^2 + \frac1{g^2\,
\sqrt{\wtd\Delta}}\, \sum_{i=1}^3 X_i^{-1}\,\Big[d\mu_i^2 +
\mu_i^2\, (d\phi_i +g\, A^i)^2\Big]\,,\label{reduction.10to5} \ee
%%%
where $X_1\, X_2\, X_3=1$.
\item For the $S^7$ reduction of $D=11$ supergravity,  the ansatz
for the eleven-dimensional metric is
%%%
\be ds_{11}^2 = \wtd\Delta^{2/3}\, ds_4^2 +g^{-2}\, \wtd\Delta^{
-1/3}\, \sum_i X_i^{-1}\, \Big[d\mu_i^2 + \mu_i^2\, (d\phi_i + g\,
A^i_{(1)})^2 \Big]\,,\label{reduction.11to4} \ee
%%%
where $\wtd\Delta = \sum_{i=1}^4 X_i\, \mu_i^2$, and $\sum_i^4
\mu_i^2 =1$ and $X_1X_2X_3X_4=1$.
\item For the $S^4$ reduction of $D=11$ supergravity, the ansatz
for the eleven-dimensional metric is
%%%
\bea ds_{11}^2&=&\wtd\Delta^{1/3}\,ds_7^2 + g^{-2}\,\wtd\Delta^{
-2/3}\, \Bigg\{X_0^{-1}\, d\mu_0^2\nn\\
&&+ \sum_{i=1}^2 X_i^{-1}\, \Big[d \mu_i^2 + \mu_i^2\, (d\phi_i +
g\, A_{(1)}^i)^2\Big] \Bigg\} \,,\label{reduction.11to7}\eea
%%%
where $\wtd\Delta = \sum_{\alpha=0}^2 X_\alpha\, \mu_\alpha^2$
with $\mu_0^2+\mu_1^2+\mu_2^2=1$, and  the auxiliary variable
$X_0\equiv (X_1 X_2)^{-2}$.
\item For the $S^4$ reduction of type IIA supergravity, the ansatz
for the ten-dimensional metric is found in
\cite{cvetic.lu.pope99},
%%%
\bea d\hat s_{10}^2&=&(\sin\xi)^{\frac1{12}}\,X^{\frac18}\Big[
\Delta^{\frac38}\,ds_6^2+2g^{-2}\,\Delta^{\frac38}\,X^2\,d\xi^2\nn\\
&&+\frac12g^{-2}\, \Delta^{-\frac58}\, X^{-1}\, \cos^2\xi
\sum_{i=1}^3(\sigma^i + g\, A_{(1)}^i)^2\Big]\,,
\label{reduction.10to6}\eea
%%%
where $X=e^{-\frac1{2\sqrt2}\phi}$, and $\Delta=X\cos^2\xi +X^{-3}
\sin^2 \xi$. The quantities $\sigma^i$ are left-invariant 1-forms
on $S^3$, which satisfy $d\sigma^i = -\frac12 \epsilon_{ijk}\,
\sigma^j\wedge \sigma^k$. One can parameterize them as
%%%
\be \sigma_1=d\theta\,,\quad \sigma_2=\sin^2\theta d\phi\,,\quad
\sigma_3=d\psi+\cos\theta d\phi\,.\ee
%%%
\end{itemize}
%%%
For all the examples that will be discussed in the following, the
lower dimension metrics will be static. So the metric will not
have any cross terms involving $d\hat t$ and the azimuthal angles.
So for the terms involved in (\ref{metric.feature1}) and
(\ref{metric.feature.def.A}), one will have $f_a=0$. What's more,
all the gauge fields are of the particular form, $A^i=\Phi^i(r)
d\hat t$; and for (\ref{reduction.10to6}), only $A_{(1)}^3\neq0$.
So it is easy to see that $h_i/h_v=-g \Phi^i(r)$. It is then
obvious that all the metrics (\ref{reduction.10to5}),
(\ref{reduction.11to4}), (\ref{reduction.11to7}), and
(\ref{reduction.10to6}) will be of the form
(\ref{metric.general.c}). Now lets look at explicit examples.

The first example is with the maximal gauged supergravity in
$D=5$. It has $SO(6)$ gauge symmetry.  The Cartan subgroup is
$U(1)^3$. The five-dimensional three-charge static AdS black hole
solution was constructed in \cite{behrndt.cvetic.sabra98}.  We
adopt the convention of \cite{10author}, and the solution is given
by
%%%
\bea ds_5^2 &=& -\cH^{-2/3} f\ d\hat t^2+\cH^{1/3}
(f^{-1} d\hat r^2+\hat r^2 d\Omega_{3,\epsilon}^2)\,,\nn\\
X_i &=& H_i^{-1} \cH^{1/3}\,, \quad A_{(1)}^i=\Phi_i\, d\hat
t\,,\quad\Phi_i = -(1-H_i^{-1}) \alpha_i \,,\nn\\
f &=& \epsilon-\frac{\mu}{\hat r^2}+g^2\hat r^2 \cH\,, \quad
\cH=H_1H_2H_3\,,\quad H_i=1+\frac{\ell_i^2}{\hat r^2}\,,\nn\\
\alpha_i &=&\frac{\sqrt{1 + \epsilon \sinh^2\beta_i}}{\sinh
\beta_i}\,, \quad \ell_i^2 = \mu\sinh^2\beta_i\,,\eea
%%%
where $d\Omega_{3,\epsilon}^2$ is the unit metric for $S^3$, $T^3$
or $H^3$ for $\epsilon=1, 0$ or $-1$, respectively.  If all the
charge parameters $\beta_i$ are set equal, the solution becomes
the five-dimensional Reissner-Nordstr\"om AdS black hole. We see
that
%%%
\bea\frac{h_i}{h_v}&=&-g\Phi_i\,,\quad h_\chi^\phi=h_{tt}=0\,,\nn\\
\cA=d\hat t+\frac{\sqrt{\cH}}{f}dr &\Longrightarrow&
h_v=\sqrt{\cH}\,,\quad f_i=0\,,\quad h_\cA=0\,.\eea
%%%

The second example is with the maximum gauged supergravity in
$D=4$. It has $SO(8)$ gauge group, with the Cartan subgroup
$U(1)^4$. The four-charge static AdS black hole was constructed in
\cite{duff.liu99,sabra99}. Following the convention of
\cite{10author}, the four-dimensional 4-charge AdS black hole
solution is given by
%%%%
\bea ds_4^2&=& -\cH^{-1/2} f\ d\hat t^2+\cH^{1/2}
(f^{-1} d\hat r^2+\hat r^2 d\Omega_{2,\epsilon}^2)\,,\nn\\
X_i &=& H_i^{-1} \cH^{1/4}\,, \qquad A_{(1)}^i=\Phi_i\, d\hat
t\,,\quad\Phi_i = -(1-H_i^{-1}) \alpha_i \,,\nn\\
f &=& \epsilon-\frac{\mu}{\hat r}+4g^2\hat r^2 \cH\,, \qquad
\cH=H_1H_2H_3H_4\,,\qquad H_i=1+\frac{\ell_i}{\hat r}\,,\nn\\
\alpha_i &=&\frac{\sqrt{1 + \epsilon \sinh^2\beta_i}}{\sinh
\beta_i}\,, \qquad \ell_i = \mu\sinh^2\beta_i\,, \eea
%%%%
where $d\Omega_{2,\epsilon}^2$ is the unit metric for $S^2$, $T^2$
or $H^2$ for $\epsilon=1, 0$ or $-1$, respectively.  If the charge
parameters $\beta_i$ are set equal, the solution becomes the
standard Reissner-Nordstr\"om AdS black hole. We see that
%%%
\bea\frac{h_i}{h_v}&=&-g\Phi_i\,,\quad h_\chi^\phi=h_{tt}=0\,,\nn\\
\cA=d\hat t+\frac{\sqrt{\cH}}{f}dr &\Longrightarrow&
h_v=\sqrt{\cH}\,,\quad f_i=0\,,\quad h_\cA=0\,.\eea
%%%

The third example is with the maximal gauged supergravity in
$D=7$. It has $SO(5)$ gauge symmetry, whose Cartan subgroup is
$U(1)^2$. The seven-dimensional 2-charge AdS black hole solution
is given by \cite{10author}
%%%%
\bea ds_7^2 &=& -\cH^{-4/5} f\ d\hat t^2+\cH^{1/5}
(f^{-1} d\hat r^2+\hat r^2 d\Omega_{5,\epsilon}^2)\,,\nn\\
X_i &=& H_i^{-1}\cH^{2/5}\,, \quad A_{(1)}^i=\Phi_i\, d\hat
t\,,\quad\Phi_i = -(1-H_i^{-1}) \alpha_i \,,\nn\\
f &=& \epsilon-\frac{\mu}{\hat r^4}+\frac14 g^2\hat r^2 \cH\,,
\quad \cH=H_1H_2\,,\quad H_i=1+\frac{\ell_i^4}{\hat r^4}\,,\nn\\
\alpha_i &=&\frac{\sqrt{1 + \epsilon \sinh^2\beta_i}}{ \sinh
\beta_i}\,, \quad \ell_i^4 = \mu\sinh^2\beta_i\,, \eea
%%%%
where $d\Omega_{5,\epsilon}^2$ is the unit metric for $S^5$, $T^5$
or $H^5$ for $\epsilon=1, 0$ or $-1$, respectively. We see that
%%%
\bea\frac{h_i}{h_v}&=&-g\Phi_i\,,\quad h_\chi^\phi=h_{tt}=0\,,\nn\\
\cA=d\hat t+\frac{\sqrt{\cH}}{f}dr &\Longrightarrow&
h_v=\sqrt{\cH}\,,\quad f_i=0\,,\quad h_\cA=0\,.\eea
%%%

The last example is with the gauged supergravity in $D=6$
constructed in \cite{romans86}. It has a $SU(2)$ gauge symmetry.
The $U(1)$ charged AdS black hole was constructed in
\cite{cvetic.lu.pope99},
%%%
\bea ds_6^2 &=& - H^{-3/2} f\, d\hat t^2 + H^{1/2} (f^{-1} d\hat
r^2 + \hat r^2 d\Omega_{4,\epsilon}^2)\,,\nn\\
X&=&H^{-1/4}\,,\quad A_{(1)} = \Phi\, d\hat t\,,\quad \Phi=
-\sqrt2(1-H^{-1}) \alpha\, d\hat t\,,\nn\\
f&=&\epsilon - \frac{\mu}{\hat r^3} + \frac29 g^2 \hat r^2
H^2\,,\quad H=1 + \frac{\ell^3}{\hat r^3}\,,\nn\\
\alpha &=& \frac{\sqrt{1 + \epsilon\,\sinh^2\beta}}{\sinh
\beta}\,,\quad \ell^3 = \mu\,\sinh^2\beta\,. \eea
%%%
We see that
%%%
\bea\frac{h_{\sigma^3}}{h_v}=-g\Phi\,,\quad h_{\sigma^1}
=h_{\sigma^2}=h_\chi^\phi=h_{tt}=0\,,\nn\\
\cA=d\hat t+\frac{H}{f}dr\quad\Longrightarrow\quad h_v=H\,,\quad
f_i=0\,,\quad h_\cA=0\,.\eea
%%%

\subsection{Extremal Rotating Black Holes in Supergravity Theories}

The Kerr/CFT correspondence for rotating black hole solutions in
supergravity theories were studied in
\cite{chow.cvetic.lu.pope08}. Here we will revisit some of the
examples by comparing them with (\ref{metric.general.c}) and
(\ref{metric.general.d}).

In the five dimensional (un)gauged supergravities, there are three
non-extremal solutions that cannot accommodate each other. They
are the three-charge two-rotation Cveti\v{c}-Youm solution
\cite{cvetic.youm96} in the ungauged supergravity, the
three-charge equal-rotation solution \cite{clp04b} and the
three-charge (two of which equal) two-rotation solution
\cite{mei.pope07} in the gauged supergravity.

The Cveti\v{c}-Youm solution is given by
%%%
\bea ds^2&=&(H_1H_2H_3)^{1/3} \left[\frac{dx^2}{4X}
+\frac{dy^2}{4Y} +\frac{U}{G}\left(d\chi -\frac{Z}{U}
d\sigma\right)^2 +\frac{X Y}{U}d\sigma^2\right]\nn\\
&&-\frac{G\left(dt+\wtd\cA\right)^2}{(H_1H_2H_3)^{2/3}}\,,\nn\\
\wtd\cA&=&\frac{2mc_1c_2c_3\left[(a^2 +b^2-y)d\sigma -a b
d\chi\right]}{x+y-2m}-\frac{2ms_1s_2s_3(a b d\sigma -y
d\chi)}{x+y}\,,\nn\\
X&=&(x+a^2)(x+b^2)-2mx\,,\quad Y=-(a^2-y)(b^2-y)\,,\nn\\
U&=&y X-x Y\,,\quad Z=a b(X+Y)\,,\quad G=(x+y)(x+y-2m)\,,\nn\\
\cA_i&=&\frac{2m}{H_i}\Big\{c_is_i dt +s_ic_jc_k \Big[a b
d\chi+(y-a^2-b^2)d\sigma\Big]\nn\\
&&\qquad\quad+c_is_js_k(a b d\sigma-y d\chi)\Big\}\,,\quad i\neq
j\neq k\,,\nn\\
X_i&=&\frac{H_1^{1/3}H_2^{1/3}H_3^{1/3}}{H_i}\,,\quad
H_i=x+y+2ms_i^2\,,\label{eg.5d.cvetic.youm}\eea
%%%
where $s_i=\sinh\delta_i\,,\, c_i=\cosh\delta_i$ and
$i,j,k=1,2,3$. The variables $\chi$ and $\sigma$ are related to
the canonical azimuthal angles by
%%%
\be\sigma=\frac{a\hat\phi_1-b\hat\phi_2}{a^2-b^2}\,,\quad
\chi=\frac{b\hat\phi_1-a\hat\phi_2}{a^2-b^2}\,.\ee
%%%
Near the horizon, $\sigma$ is playing the role of the time
direction as in the Schwarzschild solution. We have for
(\ref{metric.feature1}) and (\ref{metric.feature.def.A}),
%%%
\be\cA=d\sigma+\frac{(a^2-b^2)\sqrt{x}\,dx}{2X}\sqrt{1-\frac{y
X}{x Y}}\;.\ee
%%%
By comparing various terms, we find that
%%%
\bea h_v&=&\frac{a b(c_1^2c_2^2c_3^2 + s_1^2 s_2^2 s_3^2) -(a^2
+b^2 -2m) c_1 c_2 c_3 s_1 s_2 s_3}{a b c_1 c_2 c_3+x s_1 s_2
s_3}m\sqrt{x}\,,\nn\\
h_1&=&\frac{a(b^2+x)s_1s_2s_3-b(b^2-2m+x)c_1 c_2 c_3}{2(a b c_1
c_2 c_3+x s_1 s_2 s_3)}\sqrt{x}\,,\nn\\
h_2&=&\frac{b(a^2+x)s_1s_2s_3-a(a^2-2m+x)c_1 c_2 c_3}{2(a b c_1
c_2 c_3+x s_1 s_2 s_3)}\sqrt{x}\,,\eea
%%%
and so
%%%
\bea d\sigma&=&\frac{a}{a^2-b^2}d\hat\phi_1 -\frac{b}{a^2-b^2}
d\hat\phi_2\,,\nn\\
-\frac{U}{4Y}&=&\Big(\frac{a}{a^2-b^2}h_1 -\frac{b}{a^2-b^2}
h_2\Big)^2-\frac{y X}{4Y}\,,\nn\\
d\chi-\frac{Z}{U}d\sigma&=&-\frac{(\frac{x}{x+y} +\frac{
a^2}{x+y-2m})(a^2-y)b}{ (\frac{x y}{x+y}+\frac{a^2 b^2}{x+y
-2m})(a^2-b^2)}\Big(d\hat\phi_1-\frac{h_1}{h_v}dt\Big)\nn\\
&&+\frac{(\frac{x}{x+y} +\frac{ b^2}{x+y-2m})(b^2-y)a}{(\frac{x
y}{x+y} +\frac{a^2b^2}{x+y-2m})(a^2-b^2)}\Big(d\hat\phi_2
-\frac{h_2}{h_v}dt\Big)\nn\\
&&-\frac{(\frac{abs_1s_2s_3}{x+y-2m}-\frac{c_1c_2c_3y}{x+y})X
\sqrt{x}\;dt}{2h_v(\frac{x y}{x+y}+\frac{a^2b^2}{x+y-2m})(a b
c_1c_2c_3+s_1s_2s_3x)}\,,\nn\\
dt+\wtd\cA&=&\frac{2m(a^2-y)}{a^2-b^2}\Big(\frac{ac_1c_2c_3}{x+y
-2m}-\frac{bs_1s_2s_3}{x+y}\Big)\Big(d\hat\phi_1-\frac{h_1}{h_v}
dt\Big)\nn\\
&&+\frac{2m(b^2-y)}{a^2-b^2}\Big(\frac{as_1s_2s_3}{x+y} -\frac{b
c_1c_2c_3}{x+y-2m}\Big)\Big(d\hat\phi_2-\frac{h_2}{h_v}dt\Big)\nn\\
&&+\frac{2 m^2X\sqrt{x}\;c_1c_2c_3s_1s_2s_3 dt}{h_v(abc_1c_2c_3
+s_1s_2s_3x)(x+y-2m)(x+y)}\,.\eea
%%%
It is obvious that (\ref{eg.5d.cvetic.youm}) is of the form
(\ref{metric.general.d}) with $h_\cA,h_\chi^1,h_\chi^2\neq0$ but
$h_{tt}=0$. As a side remark, note the gauge fields can be written
as
%%%
\bea \cA_i&=&\frac{2m}{(a^2-b^2)h_i}\Bigg\{(bc_is_js_k -as_ic_j
c_k)(a^2-y)\Big(d\hat\phi_1-\frac{h_1}{h_v}dt\Big)\nn\\
&&+(bs_ic_jc_k -ac_is_j s_k)(b^2-y) \Big(d\hat\phi_2-\frac{
h_2}{h_v}dt\Big)\Bigg\}\nn\\
&&+\frac{a b c_is_i(c_j^2c_k^2 +s_j^2s_k^2)-c_jc_ks_js_k[x +c_i^2
(a^2+b^2-2m)]}{(a bc_ic_jc_k +s_is_js_k x)h_v/(m\sqrt{x})}dt\nn\\
&&+\frac{c_jc_ks_js_k X m\sqrt{x}}{(a bc_ic_jc_k +s_is_js_k
x)h_ih_v}dt\,,\quad i\neq j\neq k\,.\eea
%%%
When transforming to the coordinates on the horizon by
(\ref{metric.feature2}), only the third line will lead to a
divergence, but which can be absorbed as pure gauge.

For the three-charge equal-rotation solution in the gauged
supergravity \cite{clp04b}, the result is given by
%%%
\bea ds^2&=&R\Bigg\{-\frac{X}{f_1}dt^2+\frac{r^2}{X}dr^2
+d\theta^2+\cos^2\theta\sin^2\theta(d\phi-d\psi)^2\nn\\
&&\qquad+\frac{f_1}{R^3} \Big(\cos^2\theta d\phi +\sin^2 \theta
d\psi -\frac{f_2}{f_1}dt\Big)^2\Bigg\}\,,\nn\\
X&=&r^4-2m(r^2-\ell^2)+g^2f_1\,,\quad f_1=2m\ell^2(r^2+2m\td
s)+R^3\,,\nn\\
f_2&=&2m\ell r^2(c_1c_2c_3 -s_1s_2s_3)+4m^2\ell s_1s_2s_3\,,\nn\\
R&=&(H_1H_2H_3)^{1/3}\,,\quad H_i=r^2+2ms_i^2\;,\;i=1,2,3\,,\nn\\
\td s&=&2s_1s_2s_3(c_1c_2c_3-s_1s_2s_3)-s_1^2s_2^2-s_1^2s_3^2
-s_2^2s_3^2\,,\nn\\
\cA_i&=&\frac{2m}{h_i}\Big[c_is_idt +\ell(c_is_js_k -s_i c_j c_k)
(\cos^2\theta d\phi +\sin^2\theta d\psi)\Big]\,.
\label{eg.5d.cvetic.lu.pope}\eea
%%%
It is easy to tell that the metric is of the
(\ref{metric.general.c}) with
%%%
\be h_v=r\sqrt{f_1}\,,\quad h_\phi=h_\psi=\frac{r
f_2}{\sqrt{f_1}}\,,\quad h_\cA=h_\chi^\phi
=h_\chi^\psi=h_{tt}=0\,.\ee
%%%
After using (\ref{metric.feature2}), the gauge fields are also
regular on the horizon up to some divergence which can be absorbed
as pure gauge.

The three-charge (two of which equal) two-rotation solution in the
gauged supergravity was found in \cite{mei.pope07}, and the result
is given by
%%%
\bea ds^2&=&H_1^{2/3} H_3^{1/3} \left\{ (x^2 - y^2)
\left(\frac{dx^2}{X}-\frac{dy^2}{Y}\right) - \frac{x^2 X (dt + y^2
d\sigma)^2}{(x^2-y^2) f H_1^2}\right.\nn\\
&&\left.\qquad\qquad\quad+\frac{y^2 Y\left[dt + (x^2+2m s_1^2)
d\sigma\right]^2}{(x^2-y^2)(\gamma +y^2) H_1^2}\right.\nn\\
&&\left.- U\left(dt + y^2 d\sigma + \frac{(x^2-y^2) f H_1 \left[a
b d\sigma +(\gamma+y^2)d\chi\right]}{a b(x^2-y^2) H_3 -2m s_3 c_3
(\gamma+y^2)}\right)^2\right\}\,,\nn\\
\cA^1 &=&\cA^2= \frac{2ms_1 c_1 (dt + y^2d\sigma)}{(x^2-y^2) H_1}\,,\nn\\
\cA^3 &=&\frac{2m\left\{ s_3 c_3(dt+y^2 d\sigma)-(s_1^2-s_3^2)
\left[a b d\sigma +(\gamma+y^2) d\chi\right]\right\}
}{(x^2-y^2)H_3}\,,\label{eg.5d.mei.pope}\\
X_1 &=& X_2 = \left(\frac{H_3}{H_1}\right)^{1/3}\,,\quad
X_3=\left(\frac{H_1}{H_3}\right)^{2/3}\,,\nn\\
f&=& x^2 + \gamma + 2 m s_3^2\,,\quad \gamma = 2ab s_3 c_3 +
(a^2+b^2) s_3^2\,,\nn\\
U&=& \frac{\left[a b (x^2-y^2)H_3 - 2m s_3 c_3 (\gamma+y^2)
\right]^2}{(x^2-y^2)^2(\gamma+y^2) f H_1^2 H_3}\,,\nn\\
H_1&=&1 + \frac{2m s_1^2}{x^2-y^2}\,,\quad
H_3=1+\frac{2ms_3^2}{x^2-y^2}\,,\nn\\
X &=& \frac{-2m x^2 + (\td a^2 +x^2)(\td b^2+x^2)}{x^2}\nn\\
&&+\frac{g^2 (\td a^2 + 2m s_1^2 +x^2)(\td b^2 + 2m s_1^2 +x^2)(2m
s_3^2 +\gamma + x^2)}{x^2} \,,\nn\\
Y &=& \frac{(\td a^2+y^2)(\td b^2 + y^2)\left[1 +g^2(\gamma+y^2)
\right]}{y^2}\,,\nn\\
s_i&=&\sinh\delta_i\,,\quad c_i=\cosh\delta_i\,,\quad  \td a=a c_3
+ b s_3\,,\quad \td b=b c_3 + a s_3\,.\nn\eea
%%%
Comparing with (\ref{metric.feature1}) and
(\ref{metric.feature.def.A}), we see that
%%%
\bea\cA&=&dt + y^2 d\sigma+\frac{(x^2-y^2)\sqrt{f}H_1}{x
X}dx\nn\\
&=&dt + y^2 d\sigma+\frac{(x^2-y^2+2ms_1^2)\sqrt{f}}{x
X}dx\,,\nn\\
\Longrightarrow\quad h_v&=&\frac{(x^2+2ms_1^2)\sqrt{f}}{x}
\,,\quad h_\sigma=-\frac{\sqrt{f}}{x}\,.\eea
%%%
As a result,
%%%
\be dt + (x^2+2m s_1^2) d\sigma\propto d\sigma
-\frac{h_\sigma}{h_v}dt\,,\ee
%%%
and with $h_\chi=\ds\frac{a b+2m c_3s_3}{x\sqrt{f}}$ ,
%%%
\bea&&dt+y^2d\sigma+\frac{(x^2-y^2)f H_1\left[a b d\sigma
+(\gamma+y^2)d\chi\right]}{a b(x^2-y^2) H_3 -2m s_3 c_3
(\gamma+y^2)}\nn\\
&=&\Big\{x+2m s_1^2+\frac{(a b+2m c_3s_3)(x^2-y^2)H_1(y^2+\gamma
)}{a b(x^2-y^2)H_3-2mc_3s_3(y^2+\gamma)}\Big\}\Big(d\sigma
-\frac{h_\sigma}{h_v}dt\Big)\nn\\
&&+\frac{(y^2+\gamma)(x^2-y^2)f H_1}{a b(x^2-y^2)H_3 -2mc_3s_3
(y^2+\gamma)}\Big(d\chi-\frac{h_\chi}{h_v}dt\Big) \,.\eea
%%%
Now it is obvious that the metric in (\ref{eg.5d.mei.pope}) is of
the form (\ref{metric.general.c}). For the gauge fields, one has
%%%
\bea\cA_1&=&\cA_2=\frac{2m c_1s_1 y^2}{(x^2-y^2)H_1}\Big(d\sigma
-\frac{h_\sigma}{h_v}dt\Big)+\frac{2mc_1s_1}{x^2+2ms_1^2}dt\,,\nn\\
\cA_3&=&-\frac{2m}{(x^2-y^2)H_3}\Bigg\{\Big[a b(s_1^2-s_3^2)-c_3
s_3y^2\Big]\Big(d\sigma -\frac{h_\sigma}{h_v}dt\Big)\nn\\
&&\qquad\qquad\qquad+(s_1^2-s_3^2)(y^2+\gamma)\Big(d\chi
-\frac{h_\chi}{h_v}dt\Big)\Bigg\}\nn\\
&&+\frac{2m\Big[c_3s_3f+(a b+2m c_3s_3)(s_1^2-s_3^2)\Big]}{f
(x^2+2ms_1^2)}dt\,.\eea
%%%
Again, when (\ref{metric.feature2}) is used, the divergent pieces
can be absorbed as pure gauge.

In the following, we consider a few more solutions in dimensions
other than five. Again, all these have been studied in
\cite{chow.cvetic.lu.pope08}. We include them here just to show
the general applicability of the metric (\ref{metric.general.c})
and (\ref{metric.general.d}).

The first example is the four-charge black hole of the ungauged
supergravity in four dimension \cite{cvetic.youm96.4d,cclp04a},
%%%
\be ds_4^2=-\frac{\rho^2-2m\hat{r}}{W}\, (d{\hat t}+ B\, d{\hat
\phi})^2 +W\, \Big(\frac{d \hat{r}^2}{\Delta} + d\theta^2 +
\frac{\Delta\, \sin^2\theta\, d{\hat \phi}^2}{\rho^2-2m\hat{r}}
\Big)\,.\label{eg.4d.four.charge}\ee
%%%
The detail of various functions can be found in
\cite{chow.cvetic.lu.pope08}. Notably,
%%%
\bea \Delta &=& \hat{r}^2 -2 m \hat{r} + a^2\,,\quad \rho^2=
\hat{r}^2+a^2\cos^2\theta\,,\quad W=W(r)\,,\nn\\
B &=& \frac{2ma^2\sin^2\theta[\hat{r}c_1c_2c_3c_4- (\hat{r}-2m)
s_1s_2s_3s_4]}{a(\rho^2 -2m\hat{r})}\,.\eea
%%%
Note $\rho^2 -2m\hat{r}=\Delta-a^2\sin^2\theta$. So when it comes
close to the horizon, $d\hat\phi$ replaces $d{\hat t}+ B\, d{\hat
\phi}$ and become the time direction. What's more,
%%%
\bea B&=&-\frac1{B_0}\Big(1+\frac{\Delta}{a^2\sin^2\theta}\Big)
+\cO(\Delta^2)\,,\nn\\
B_0&=&\frac{a}{2m[\hat{r}c_1c_2c_3c_4-(\hat{r}-2m)s_1 s_2s_3
s_4]}\,.\eea
%%%
Comparing (\ref{eg.4d.four.charge}) with (\ref{metric.feature1}),
we have for (\ref{metric.feature.def.A}),
%%%
\bea\cA&=&d\hat\phi+\frac{\sqrt{a^2\sin^2\theta-\Delta}}{\Delta
\sin\theta}d\hat r\nn\\
&\approx&d\hat\phi+\frac{a}{\Delta}d\hat r -\frac{d\hat
r}{2a\sin^2\theta}\,,\nn\\
\Longrightarrow\quad h_{\hat\phi}&=&a\,,\quad
h_\cA=-\frac{1}{2a\sin^2\theta}\,.\eea
%%%
By letting $h_v=\ds\frac{a}{B_0}$ and $h_\chi^{\hat\phi}
=-\ds\frac1{a\sin^2\theta}$ , we also have
%%%
\be d{\hat t}+ B\, d{\hat \phi}\propto d\hat\phi-\frac{h_{\hat
\phi} +h_\chi^{\hat\phi}\Delta}{h_v}d\hat t+\cO(\Delta^2)\,.\ee
%%%
So (\ref{eg.4d.four.charge}) is of the form
(\ref{metric.general.d}) with $h_{tt}=0$.

The next example is the rotating black hole solution in
four-dimensional U(1)$^4$ gauged supergravity with the four U(1)
charges pairwise equal \cite{cclp04a}. The metric is
%%%
\bea ds^2&=&H\Big[-\frac{R}{H^2(\hat{r}^2 + y^2)}\left(d\hat{t}
-\frac{a^2-y^2}{\Xi a}d\hat{\phi}\right)^2+\frac{\hat{r}^2
+y^2}{R}d\hat{r}^2+\frac{\hat{r}^2+y^2}{Y}dy^2\nn\\
&&+\frac{Y}{H^2(\hat{r}^2+y^2)}\left(d\hat{t}-\frac{(\hat{r}+q_1)
(\hat{r}+q_2)+a^2}{\Xi a}d\hat{\phi}\right)^2\Big]\,,
\label{eg.4d.four.charge.pair.equal}\eea
%%%
where
%%%
\bea R&=&\hat{r}^2+a^2+g^2(\hat{r}+q_1)(\hat{r}+q_2)[(\hat{r}+
q_1)(\hat{r}+q_2)+a^2]-2m\hat{r}\,,\nn\\
Y&=&(1-g^2y^2)(a^2-y^2)\,,\quad\Xi=1-g^2a^2\,,\nn\\
H&=&\frac{(\hat{r}+q_1)(\hat{r}+q_2)+y^2}{\hat{r}^2+y^2}\,,\quad
q_I=2ms_I^2\,,\quad s_I=\sinh\delta_I\,. \eea
%%%
Comparing (\ref{eg.4d.four.charge.pair.equal}) with
(\ref{metric.feature1}), we have for (\ref{metric.feature.def.A}),
%%%
\bea\cA&=&d\hat{t} -\frac{a^2-y^2}{\Xi a}d\hat{\phi} +\frac{
(\hat{r}+q_1)(\hat{r}+q_2)+y^2}{R}d\hat r\,,\nn\\
\Longrightarrow\quad h_v&=&\frac{ (\hat{r}+q_1)(\hat{r}+q_2)
+a^2}{R}\,,\quad h_{\hat\phi}=\frac{\Xi a}{R}\,.\eea
%%%
It is easy to see that
%%%
\be d\hat{t}-\frac{(\hat{r}+q_1) (\hat{r}+q_2)+a^2}{\Xi a} d\hat
\phi\quad\propto\quad d\hat\phi-\frac{h_{\hat\phi}}{h_v} d\hat
t\,.\ee
%%%
So (\ref{eg.4d.four.charge.pair.equal}) is of the form
(\ref{metric.general.c}) with $h_\cA=h_\chi^{\hat\phi} =h_{tt}=0$.

A single-charge two-rotation solution to  the six-dimensional
SU(2) gauged supergravity was found in \cite{chow08}. The metric
is
%%%
\bea d s^2&=&H^{1/2}\Bigg\{-\frac{R}{H^2U}\wtd\cA^2
+\frac{(\hat{r}^2+y^2)(y^2-z^2)}{Y}dy^2
+\frac{Y\wtd\cA_Y^2}{(\hat{r}^2+y^2)(y^2-z^2)}\nn\\
&&\qquad\qquad+\frac{U}{R} d\hat{r}^2
+\frac{(\hat{r}^2+z^2)(z^2-y^2)}{Z}dz^2 +\frac{Z\wtd
\cA_Z^2}{(\hat{r}^2+z^2)(z^2-y^2)}\Bigg\}\,,\nn\\
\label{eg.6d.single.charge}\\
\wtd\cA_Y&=&d\hat{t}-(\hat{r}^2+a^2)(a^2-z^2) \frac{d\hat{
\phi}_1}{\epsilon_1}-(\hat{r}^2+b^2)(b^2-z^2) \frac{d\hat{
\phi}_2}{\epsilon_2}-\frac{q\hat{r}\wtd\cA}{H U}\,,\nn\\
\wtd\cA_Z&=&d\hat{t}-(\hat{r}^2+a^2)(a^2-y^2)\frac{d\hat{
\phi}_1}{\epsilon_1}-(\hat{r}^2+b^2)(b^2-y^2) \frac{d\hat{
\phi}_2}{\epsilon_2}-\frac{q\hat{r}\wtd\cA}{H U}\,,\eea
%%%
where the various functions and constants can be found in
\cite{chow.cvetic.lu.pope08}. The ones relevant for us are
%%%
\bea U&=&(\hat{r}^2+y^2)(\hat{r}^2 + z^2)\,,\quad H=1+\frac{q
\hat{r}}{U}\,,\nn\\
\wtd\cA&=&d\hat{t}-(a^2-y^2)(a^2-z^2)\frac{d\hat{\phi}_1}{\epsilon_1}
-(b^2-y^2)(b^2-z^2)\frac{d\hat{\phi}_2}{\epsilon_2}\,.\eea
%%%
Comparing (\ref{eg.6d.single.charge}) with
(\ref{metric.feature1}), we have for (\ref{metric.feature.def.A}),
%%%
\be\cA=\wtd\cA+\frac{H U}{R}dr\,.\ee
%%%
By comparing various terms, one can find
%%%
\bea h_v&=&(\hat r^2+a^2)(\hat r^2+b^2)+q\hat r\,,\nn\\
h_1&=&\frac{\hat r^2+b^2}{a^2-b^2}\epsilon_1\,,\quad
h_2=\frac{\hat r^2+a^2}{b^2-a^2}\epsilon_2\,,\eea
%%%
and
%%%
\bea\wtd\cA_Y&=&\frac{(z^2-a^2)[q\hat r+(\hat r^2+a^2)(\hat r^2
+z^2)](\hat r^2+y^2)}{H U \epsilon_1}\Big(d\hat\phi_1
-\frac{h_1}{h_v}d\hat t\Big)\nn\\
&&+\frac{(z^2-b^2)[q\hat r+(\hat r^2+b^2)(\hat r^2+z^2)](\hat r^2
+y^2)}{H U \epsilon_2}\Big(d\hat\phi_2 -\frac{h_2}{h_v}d\hat
t\Big)\,,\nn\\
\wtd\cA_Z&=&\frac{(y^2-a^2)[q\hat r+(\hat r^2+a^2)(\hat r^2
+y^2)](\hat r^2+z^2)}{H U \epsilon_1}\Big(d\hat\phi_1
-\frac{h_1}{h_v}d\hat t\Big)\nn\\
&&+\frac{(y^2-b^2)[q\hat r+(\hat r^2+b^2)(\hat r^2+y^2)](\hat r^2
+z^2)}{H U \epsilon_2}\Big(d\hat\phi_2 -\frac{h_2}{h_v}d\hat
t\Big)\,.\eea
%%%
So (\ref{eg.6d.single.charge}) is of the form
(\ref{metric.general.c}) with $h_\cA=h_\chi^{\hat\phi} =h_{tt}=0$.

The single-charge three-rotation black hole solution to the
seven-dimensional SO(5) gauged supergravity was found in
\cite{chow07}. The metric is
%%%
\bea d s^2&=&H^{2/5}\Bigg\{-\frac{R}{H^2 U}\wtd\cA^2+\frac{U}{R}
d\hat{r}^2 +\frac{(\hat r^2 +y^2)(y^2-z^2)}{Y}dy^2\nn\\
&&\qquad+\frac{(\hat{r}^2 +z^2)(z^2-y^2)}{Z}dz^2+\frac{Y\wtd
\cA_Y^2}{(\hat{r}^2 +y^2)(y^2-z^2)}\nn\\
&&\qquad+\frac{Z\wtd\cA_Z^2}{(\hat{r}^2+z^2)(z^2-y^2)}+\frac{
a_1^2a_2^2 a_3^2}{\hat{r}^2 y^2 z^2}\wtd\cA_7^2\Bigg\}\,,\nn\\
\wtd\cA_Y&=&d\hat{t} - \sum_{i=1}^3 \frac{(\hat{r}^2 + a_i^2)
\gamma_i}{a_i^2 - y^2} \frac{d \hat{\phi}_i}{\epsilon_i} -
\frac{q}{H U} \wtd\cA\,,\nn\\
\wtd\cA_Z&=&d \hat{t} - \sum_{i=1}^3 \frac{(\hat{r}^2 + a_i^2)
\gamma_i}{a_i^2 - z^2} \frac{d \hat{\phi}_i}{\epsilon_i} -
\frac{q}{H U} \wtd\cA\,,\nn\\
\wtd\cA_7&=&d \hat{t} - \sum_{i=1}^3 \frac{(\hat{r}^2 + a_i^2)
\gamma_i}{a_i^2} \frac{d \hat{\phi}_i}{\epsilon_i} - \frac{q}{H U}
\left( 1 + \frac{g y^2 z^2}{a_1 a_2 a_3} \right)\wtd\cA\,,
\label{eg.7d.single.charge}\eea
%%%
where the various functions and constants can be found in
\cite{chow.cvetic.lu.pope08}. The ones relevant for us are
%%%
\bea U&=&(\hat{r}^2+y^2)(\hat{r}^2+z^2)\,,\quad
\gamma_i=a_i^2(a_i^2-y^2)(a_i^2-z^2)\,,\nn\\
H&=&1+\frac{q}{(\hat{r}^2+y^2)(\hat{r}^2+z^2)}\,,\quad
\wtd\cA=d\hat{t}-\sum_{i=1}^3\gamma_i\frac{d \hat
\phi_i}{\epsilon_i}\,.\eea
%%%
Comparing (\ref{eg.7d.single.charge}) with
(\ref{metric.feature1}), we have for (\ref{metric.feature.def.A}),
%%%
\be\cA=\wtd\cA+\frac{H U}{R}dr\,.\ee
%%%
By comparing various terms, one can find
%%%
\bea h_v&=&\frac{(r^2+a_1^2)(r^2+a_2^2)(r^2+a_3^2)+q(r^2-ga_1
a_2a_3)}{r^2}\,,\nn\\
h_i&=&\frac{a_i(r^2+a_j^2)(r^2+a_k^2)-gqa_ja_k}{a_i(a_i^2-a_j^2)
(a_i^2-a_k^2)r^2}\epsilon_i\,,\quad i\neq j\neq k\,,\eea
%%%
and
%%%
\bea\wtd\cA_Y&=&\sum_{i=1}^3\frac{(z^2-a_i^2)[q+(\hat r^2 +a_i^2)
(\hat r^2 +z^2)](\hat r^2+y^2)a_i^2}{H U \epsilon_i} \Big(d\hat
\phi_i-\frac{h_i}{h_v}d\hat t\Big)\,,\nn\\
\wtd\cA_Z&=&\sum_{i=1}^3\frac{(y^2-a_i^2)[q+(\hat r^2 +a_i^2)
(\hat r^2 +y^2)](\hat r^2+z^2)a_i^2}{H U \epsilon_i} \Big(d\hat
\phi_i-\frac{h_i}{h_v}d\hat t\Big)\,,\nn\\
\wtd\cA_7&=&\sum_{i=1}^3\frac{\gamma_i\Big[\ds\frac{q(a_1a_2a_3+g
y^2 z^2)}{H U}-\frac{a_1a_2a_3}{a_i^2}(r^2+a_i^2)\Big]}{a_1a_2a_3
\epsilon_i}\Big(d\hat\phi_i-\frac{h_i}{h_v}d\hat t\Big)\,.\eea
%%%
So (\ref{eg.7d.single.charge}) is of the form
(\ref{metric.general.c}) with $h_\cA=h_\chi^{\hat\phi} =h_{tt}=0$.

%%%%%%%%%%%%%%%%%%%%%%%%%%%%%%%%%%%
\end{document}